\documentclass[pre,showpacs,floatfix,onecolumn,superscriptaddress]{revtex4}

\usepackage{graphicx}
\usepackage{subfigure}
\usepackage{amsmath}
\usepackage{amssymb}
\usepackage{latexsym}
\usepackage{multirow}
\usepackage{color}
\usepackage[hidelinks]{hyperref}
\usepackage{mathtools}
\usepackage{comment}

\renewcommand{\Re}{\operatorname{Re}}

\newcommand{\etal}{{\sl et al. }}

\begin{document}

\title{Flocking with a $q$-fold discrete symmetry:\\band-to-lane transition in the active Potts model}

\author{Matthieu Mangeat}
\email{mangeat@lusi.uni-sb.de}
\affiliation{Center for Biophysics \& Department for Theoretical Physics, Saarland University, D-66123 Saarbr{\"u}cken, Germany.}

\author{Swarnajit Chatterjee}
\email{sspsc5@iacs.res.in}
\affiliation{School of Mathematical \& Computational Sciences, Indian Association for the Cultivation of Science, Kolkata -- 700032, India.}

\author{Raja Paul}
\email{raja.paul@iacs.res.in}
\affiliation{School of Mathematical \& Computational Sciences, Indian Association for the Cultivation of Science, Kolkata -- 700032, India.}

\author{Heiko Rieger}
\email{h.rieger@mx.uni-saarland.de}
\affiliation{Center for Biophysics \& Department for Theoretical Physics, Saarland University, D-66123 Saarbr{\"u}cken, Germany.}

\begin{abstract}
We study the $q$-state active Potts model (APM) on a two-dimensional lattice in which self-propelled particles have $q$ internal states corresponding to the $q$ directions of motion. A local alignment rule inspired by the ferromagnetic $q$-state Potts model and self-propulsion via biased diffusion according to the internal particle states leads to a collective motion at high densities and low noise. We formulate a coarse-grained hydrodynamic theory of the model, compute the phase diagrams of the APM for $q=4$ and $q=6$ and explore the flocking dynamics in the region, in which the high-density (polar liquid) phase coexists with the low-density (gas) phase and forms a fluctuating stripe of coherently moving particles. As a function of the particle self-propulsion velocity, a novel reorientation transition of the phase-separated profiles from transversal band motion to longitudinal lane formation is revealed, which is absent in the Vicsek model and APM for $q=2$ (active Ising model). The origin of this reorientation transition is obtained via a stability analysis: for large velocities, the transverse diffusion constant approaches zero and stabilizes lanes. Finally, we perform microscopic simulations that corroborate our analytical predictions about the flocking and reorientation transitions and validate the phase diagrams of the APM.
\end{abstract}

\maketitle


\section{Introduction}
\label{section1}

Active matter systems are natural or synthetic systems composed of large numbers of particles that consume energy in order to move or to exert mechanical forces. An assembly of active particles displays a complex dynamics and collective effects like the emergence of ordered motion of large clusters, called flocks, with a typical size larger than one individual \cite{ramaswamy,shaebani,magistris,menon}. Flocking plays a significant role in a wide range of systems across disciplines including physics, biology, ecology, social sciences, and neurosciences \cite{strogatz} and is an out of equilibrium phenomenon abundantly observed in nature \cite{marchetti2013}: from human crowds \cite{bottinelli2016,helbing1995}, mammalian herds \cite{garcimartin2015}, bird flocks \cite{ballerini2008}, fish schools \cite{beccoa2006,calovi2014} to unicellular organisms such as amoebae, bacteria \cite{steager2008,peruani2012}, collective cell migration in dense tissues \cite{giavazzi2018}, and sub-cellular structures including cytoskeletal filaments and molecular motors \cite{schaller2010,sumino2012,sanchez2012}. Collective dynamics is also prevalent in nonliving systems such as rods on a horizontal surface agitated vertically \cite{deseigne, weber2013}, self-propelled liquid droplets \cite{shashi2011} and rolling colloids \cite{bricard2013}. Despite the huge differences in the scales of aggregations for different active matter systems, the similarities in the patterns suggested that there might be a general principle of flocking.

A widely studied computational model for flocking is due to Vicsek and coworkers \cite{Viscek}. In this model, an individual particle tends to align with the average direction of the motion of its neighbors. At low noise and high density, the particles cluster and move collectively in a common direction, which is the characteristic of flocking. Toner and Tu \cite{toner} developed a theoretical model describing a large universality class of active systems, including the Vicsek model (VM). They have convincingly shown that the coherent motion of the flock is a phase with spontaneously broken symmetry with no preferred direction, each flock spontaneously selects an arbitrary direction to move.

Due to the rich physics of the VM \cite{ginelli}, numerous analytical and computational studies were carried out by several research groups contributing significantly to the understanding of the principles of the flocking transition. Two ingredients are important: the interactions between the particles (alignment and/or repulsion) and the shape of the particle. For the Vicsek-like models, pointlike polar particles align with ferromagnetic interactions having no repulsion \cite{GC2004, TT2005, bertin2009, marchetti, ihle, solon2015, liebchen2017, sandor2017, escaff2018, miguel2018}. In the VM a region in the noise-density phase diagram a region exists in which the disordered phase and the ordered (flocked) phase coexist. In this coexistence region, the ordered phase forms stripes moving perpendicularly to the average motion direction of the particles, so-called bands. These bands have a maximum width, implying the formation of several bands in a large system, which is denoted as {\it microphase} separation \cite{solon2015}. The transverse orientation of the stripes emerging in the VM has been understood within a hydrodynamic theory \cite{bertin2009}, which predicts that the long-wavelength instability is stronger in the longitudinal direction. In the presence of repulsive interactions, in addition to the local alignment rule \cite{peruani2011, farrell2012, MG2018}, more patterns of collective motion emerge such as asters (immobile clusters), bands (transverse stripes) and lanes (longitudinal stripes).

In addition to ferromagnetic (Vicsek-like) alignment interactions, nematic alignment between particles has been studied as well \cite{peshkov,julicher2018}. Examples are self-propelled elongated particles with excluded volume interactions, either self-propelled rods (polar particles) \cite{marchetti2008, ginelli2010} or active nematics (apolar particles) \cite{chate2006, bertin2013, bertin2014}. In these systems with nematic alignment, stripe formation can be observed, nevertheless there the long-wavelength instability is stronger in the perpendicular direction with respect to the collective motion giving rise to lanes.

Further insights into the flocking transitions gave recent studies on the active Ising model (AIM) \cite{ST2013,ST2015,ST2015-2}. Here it was argued that the flocking transition can be seen as a liquid-gas phase-separation rather than an order-disorder transition. Upon increasing the density at low noise the system undergoes a transition from a disordered gaseous phase to an ordered liquid phase with an intermediate liquid-gas coexistence phase. The continuous symmetry of the Vicsek model has been replaced by discrete symmetry in the AIM, nevertheless, the model contained the rich physics of the flocking transition in a much simpler and tractable manner. The main difference with the VM lies in the full phase-separation of the bands.

In this paper, we study the generalization of the AIM: the $q$-state Active Potts Model (APM), which involves $q$ internal spin-states: $q=2$ corresponds to the earlier AIM and $q=4$ was previously studied in Ref.~\cite{chatterjee2020}. We consider the APM on two-dimensional lattices with coordination number $q$, for instance, a square lattice for $q=4$ and a triangular lattice for $q=6$. The two main ingredients for flocking are local ferromagnetic alignment between the on-site particles inspired by the standard $q$-state Potts model and self-propulsion via biased hopping to the nearest-neighbor lattice sites in one of the $q$ possible directions given by the spin-state, without repulsive interactions. We determine the stationary state of this model, displaying the collective motion in large regions of the phase diagram, by constructing a coarse-grained hydrodynamic theory and analyzing the microscopic Monte Carlo simulations. The main findings of our study are: (i) The flocking transition in the APM is a liquid-gas phase transition as observed in the AIM subject to temperature $T=\beta^{-1}$, average particle density $\rho_0$, and self-propulsion velocity $\epsilon$. (ii) In the co-existence phase, the liquid domains coalesce to form a stripe-like structure, oriented transversely (denoted in the following as a {\it band}) or longitudinally (denoted in the following as a {\it lane}) and moving perpendicular or parallel to the internal spin-state governing the liquid phase of the stripes. This property leads to a novel reorientation transition, depending upon the self-propulsion velocity $\epsilon$, absent in the Vicsek model and AIM. (iii) The characterization of the $\epsilon=0$ critical point as a first-order phase transition from high density ordered phase to low density disordered phase for $q$-state APM ($q \ge 4$), different from the standard $q$-state Potts model.

We start this paper by defining the microscopic model and technical details of the simulation protocols in Sec.~\ref{section2}. From the microscopic rules for the ferromagnetic alignment and the hopping to neighboring lattice sites, we construct a coarse-grained hydrodynamic theory in Sec.~\ref{section3}. By solving the hydrodynamic equations for the spatio-temporal evolution of the particle densities corresponding to the $q$ different directions of motion numerically we determine the stationary states, construct the corresponding phase diagrams and analyze the flocking transition and the novel reorientation transition. In Sec.~\ref{section4}, the homogeneous solutions of these PDEs are derived analytically. Then, we perform a linear stability analysis which reveals the physical origin of the reorientation transition. In Sec.~\ref{section5}, we present the results of extensive Monte Carlo simulations of the microscopic model and compare the resulting phase diagrams with those predicted by the hydrodynamic theory. Finally, we conclude this paper with a summary and discussion of the results in Sec.~\ref{section6}.


\section{Model}
\label{section2}
\begin{figure}[t]
\centering
\includegraphics[width=16cm]{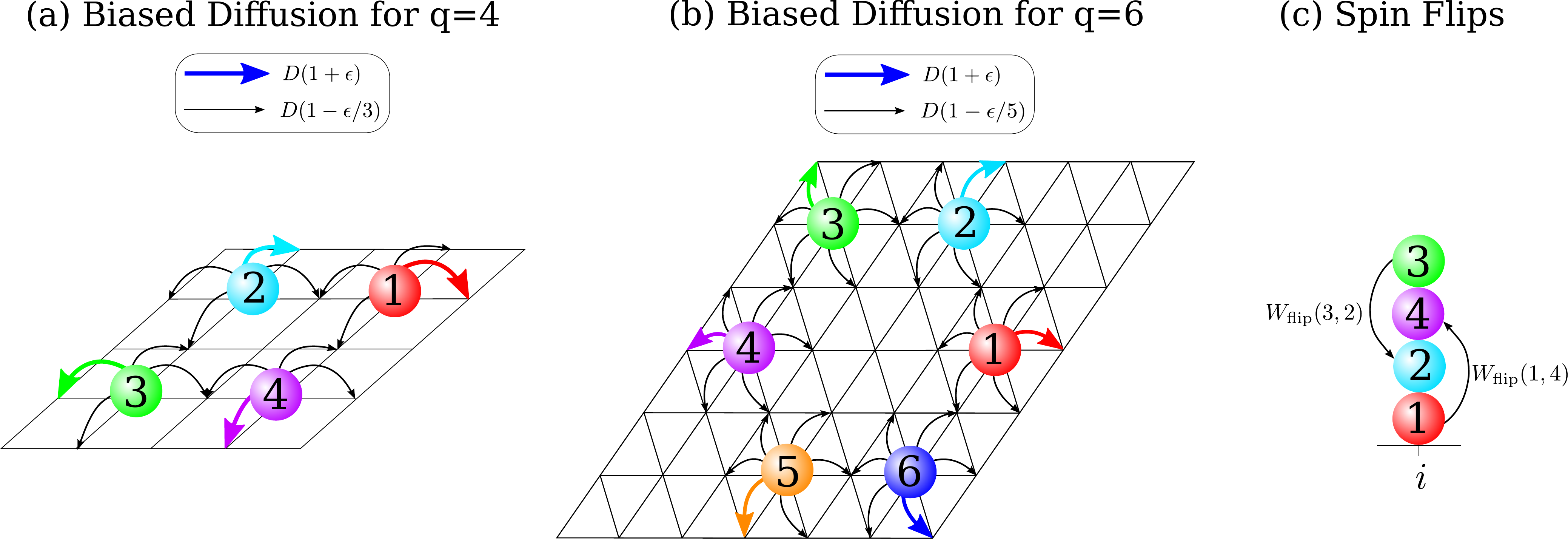}
\caption{(color online) Schematic of APM showing biased hopping rates to a nearest neighbor lattice site for (a) $q=4$ and (b) $q=6$ states, with rates $D(1+\epsilon)$ in the favored direction and $D[1-\epsilon/(q-1)]$ in other directions (more than one particle is allowed on a site but not shown in (a) and (b) due to clarity of representation). (c) Spin flips (i.e. directional changes) are performed locally (i.e. without moving the particle) with a rate $W_{\rm flip}(\sigma,\sigma')$ for a particle changing direction from state $\sigma$ to state $\sigma'$.}
\label{fig0}
\end{figure}

We consider an ensemble of $N$ particles defined on two-dimensional lattices (e.g. square or triangular) of linear size $L$ with periodic boundary conditions. The average particle density is $\rho_0 = N/L^2$. No restriction is applied on the number of particles on a given lattice site and a particle on site $i$ with a given spin-state $\sigma$ can either flip to a different spin-state $\sigma'$ or jump to a nearest neighbor site probabilistically. A schematic diagram of this arrangement is shown in Fig.~\ref{fig0}. The spin-state of the $k$-th particle on lattice site $i$ is denoted $\sigma_i^k$, with an integer value in $[1, q]$, while the number of particles in state $\sigma$ on site $i$ is $n_i^\sigma$. The local density on site $i$ is then defined by $\rho_i=\sum_{\sigma=1}^{q} n_i^{\sigma}$, counting the total number of particles on the site. The flip probabilities are derived from a ferromagnetic Potts Hamiltonian $H_{\rm APM}= \sum_i H_i$ decomposed as the sum of local Hamiltonians $H_i$:
\begin{equation}
\label{Hpotts}
H_i = - \frac{J}{2\rho_i}\sum_{k=1}^{\rho_i}\sum_{l \neq k}(q\delta_{\sigma_i^k,\sigma_i^l}-1),
\end{equation}
where $J$ is the coupling between the particles on site $i$ and the Kronecker delta $\delta_{\sigma_i^k,\sigma_i^l}$ survives only for $\sigma_i^k = \sigma_i^l$. Eq.~(\ref{Hpotts}) with $q=2$ is equivalent to the local Hamiltonian defined for the AIM \cite{ST2015}. The local magnetization corresponding to state ${\sigma}$ on site $i$ is defined as
\begin{equation}
\label{mag}
m_i^{\sigma}=\sum_{k=1}^{\rho_i}\frac{q\delta_{\sigma,\sigma_i^k}-1}{q-1} = \frac{q n_i^\sigma - \rho_i}{q-1}.
\end{equation}
Eq.~(\ref{mag}) with $q=2$ retrieves the expression of local magnetization for the AIM \cite{ST2015}.

The spin flip transition rates are derived form the Potts Hamiltonian $H_{\rm APM}$ according to the energy difference between the new and the old state. Consider a spin flip of a single particle on site $i$ from state $\sigma$ to state $\sigma'$. Since only the on-site energy is changed we rewrite the on-site Hamiltonian as
\begin{equation}
H_i = - \frac{qJ}{2\rho_i} \sum_{\sigma=1}^{q} {n_i^\sigma}^2 + \frac{J}{2} (\rho_i + q - 1).
\end{equation}
Then the energy difference between the new and the old state is
\begin{equation}
\Delta H = H_i^{\rm new}-H_i^{\rm old} = - \frac{qJ}{2\rho_i} \left[ (n_i^\sigma-1)^2 + (n_i^{\sigma'}+1)^2 \right] + \frac{qJ}{2\rho_i} \left[ {n_i^\sigma}^2 + {n_i^{\sigma'}}^2 \right] = \frac{qJ}{\rho_i}\left(n_i^{\sigma}-n_i^{\sigma'}-1 \right).
\end{equation}
Then, in analogy to the AIM \cite{ST2015}, the transition rate is chosen in such a way that without hopping detailed balance with respect to the Hamiltonian $H_{\rm APM}$ would be fulfilled, i.e. for a flip of $k$-th particle on site $i$ from state $\sigma$ to state $\sigma'$
\begin{equation}
\label{defWflip}
W_{\rm flip}(\sigma,\sigma')= \gamma \exp \left[-\frac{q\beta J}{\rho_i} (n_i^\sigma - n_i^{\sigma'}-1)\right].
\end{equation}

Moreover, each particle performs a biased diffusion on the lattice depending on the particle state $\sigma$: a particle in state $\sigma$ prefers to hop in the direction connected to its state $\sigma$. Evidently, on a lattice, the number of nearest neighbors should be equal to the number of states that a particle can assume. The hopping rate of a particle with state $\sigma$ in the direction $p$ is defined as
\begin{equation}
\label{defWhop}
W_{\rm hop}(\sigma,p)=D \left[1+ \frac{q\delta_{\sigma, p}-1}{q-1} \ \epsilon \right].
\end{equation}
$\epsilon$ parametrizes the strength of the self-propulsion \cite{ST2015}, i.e. $\epsilon=0$ describes purely diffusive particles and $\epsilon=q-1$ describes purely ballistic particles, implying $0 \le \epsilon \le q-1$. When $q=2$, the maximum value of $\epsilon$ is 1, which defines a fully self-propelled dynamics in the AIM \cite{ST2015}. According to Eq.~(\ref{defWhop}), the hopping rate is $W_{\rm hop}=D(1+\epsilon)$, when $\sigma=p$ and $W_{\rm hop}=D[1-\epsilon/(q-1)]$, otherwise. Note that the total hopping rate is $qD$, and does not depend on $\epsilon$.

In the limit $q\to\infty$, the total hopping rate $q D$, the maximal value of $\epsilon$ and the total flipping rate must stay finite. Therefore in this limit one has to rescale the microscopic parameters such that $\overline D = q D$, $\overline \epsilon = \epsilon/(q-1)$ and $\overline \gamma = q \gamma \exp(q \beta J/\rho_0)$ are independent of $q$. Each particle has then a continuous spin-state pointing along direction $\phi \in [0,2\pi]$ and the density of particles $n_i(\phi)$ in the state $\phi$ replaces the number of particles $n_i^\sigma$ in the state $\sigma$ which approaches zero in the limit $q\to\infty$ (see also appendix \ref{appendixInfty}). The hopping rates, Eq.~(\ref{defWhop}), are such that a particle in state $\phi$ jumps in a random direction with rate $\overline D(1- \overline \epsilon)$ or in the direction $\phi$ with rate $\overline D \overline \epsilon$. The flipping rate, Eq.~(\ref{defWflip}), is replaced by a flipping rate density for a spin flip from state $\phi$ to state $\phi'$:
\begin{equation}
W_{\rm flip}(\phi,\phi') = \frac{\overline \gamma}{2\pi} \exp\left\{-\frac{2\pi \beta J}{\rho_i} \left[n_i(\phi) - n_i(\phi')\right]\right\}.
\end{equation}
Since the new direction of motion after the flip is not necessarily close to $\phi$, the Vicsek model is not recovered in this $q\to\infty$ limit. In the disordered phase (at high temperatures) the particles perform a run-and-tumble process \cite{berg1972,rieger2019} with a bias depending on $\epsilon$, whereas in the ordered phase (at low temperatures) the particles flock together and move in the same direction. 

The stochastic process defined with the transition rates of Eqs.~(\ref{defWflip}) and (\ref{defWhop}) can be realized with Monte Carlo simulation in discrete time steps $\Delta t$ in which $N$ single particle spin flips or hopping events are generated and accepted according to the given rates \cite{ST2015}. In the microscopic time interval $\Delta t/N$ a randomly chosen particle either updates its spin state $\sigma$ to the new state (direction) $\sigma'$, chosen among the possible $q-1$ states, with probability $p_{\rm flip}=W_{\rm flip}(\sigma, \sigma')\Delta t$ or hops to one of the neighboring sites with probability $p_{\rm hop}=\sum_{p=1}^q W_{\rm hop}(\sigma,p) \Delta t= q D \Delta t$. The probability that nothing happens is represented by $p_{\rm wait}=1-(p_{\rm flip}+p_{\rm hop})$. The sum of probabilities $p_{\rm flip}$ and $p_{\rm hop}$ is
\begin{equation}
\label{prob_sum}
p_{\rm flip}+p_{\rm hop}=\left\{qD+\gamma \exp\left[-\frac{q\beta J}{\rho_i}\left(n_i^{\sigma}-n_i^{\sigma'}-1\right)\right]\right\} \Delta t < \left[qD+\gamma \exp(q\beta J)\right] \Delta t.
\end{equation}
$p_{\rm wait}$ will be minimum when the sum of $p_{\rm flip}$ and $p_{\rm hop}$ is maximum. Now, to keep $p_{\rm wait}$ positive, the quantity $p_{\rm flip}+p_{\rm hop}$ must be smaller than $1$ and implying the inequality $\Delta t \le [qD+\gamma \exp(q\beta J)]^{-1}$. We take the largest time interval possible
\begin{equation}
\Delta t = [qD+\gamma \exp(q\beta J)]^{-1},
\end{equation}
to reduce $p_{\rm wait}$ to a minimum and thus save computation time.


\section{Hydrodynamic equations and their numerical solutions}
\label{section3}

\subsection{Derivation of hydrodynamic equations}

In this section, we will derive the hydrodynamic equation for the $q$-state APM. From the flipping and hopping rules, we can derive the master equation defining the dynamic equation for the number of particles $n_i^\sigma(t)$ on site $i$ in state $\sigma$:
\begin{align}
\langle n_i^\sigma(t+dt) \rangle &= \left\langle n_i^\sigma(t) \left[ 1 - dt \sum_{p=1}^q W_{\rm hop}(\sigma,p) - dt \sum_{\sigma' \ne \sigma } W_{\rm flip}(\sigma,\sigma') \right] \right\rangle \nonumber\\
&+ \left\langle \sum_{p=1}^q n_{i-p}^\sigma(t) W_{\rm hop}(\sigma,p) dt + \sum_{\sigma' \ne \sigma } n_i^{\sigma'}(t) W_{\rm flip}(\sigma',\sigma) dt \right\rangle,
\end{align}
where the subscript $i+p$ denotes the neighbor of site $i$ in the $p$-direction. In the limit $dt \rightarrow 0$, this expression yields
\begin{equation}
\label{MasterEq0}
\partial_t \langle n_i^\sigma \rangle = \sum_{p=1}^q W_{\rm hop}(\sigma,p) \left[ \langle n_{i-p}^\sigma \rangle - \langle n_i^\sigma \rangle \right] + \sum_{\sigma' \ne \sigma } \left\langle n_i^{\sigma'} W_{\rm flip}(\sigma',\sigma) - n_i^\sigma W_{\rm flip}(\sigma,\sigma') \right\rangle
\end{equation}
With Eq.~(\ref{defWhop}) for the hopping rates $W_{\rm hop}$ can be expressed as
\begin{equation}
\sum_{p=1}^q W_{\rm hop}(\sigma,p) \left[ \langle n_{i-p}^\sigma \rangle - \langle n_i^\sigma \rangle \right] = D \left(1 - \frac{\epsilon}{q-1} \right) \sum_{p=1}^q \left[ \langle n_{i-p}^\sigma \rangle - \langle n_i^\sigma \rangle \right] + \frac{q D \epsilon}{q-1} \left[ \langle n_{i-\sigma}^\sigma \rangle - \langle n_i^\sigma \rangle \right],
\end{equation}
where the first term corresponds to a jump in a random direction and the second term to a jump in the favored direction. With this the master equation (\ref{MasterEq0}) is
\begin{equation}
\label{MasterEq}
\partial_t \langle n_i^\sigma \rangle = D \left(1 - \frac{\epsilon}{q-1} \right) \sum_{p=1}^q \left[ \langle n_{i-p}^\sigma \rangle - \langle n_i^\sigma \rangle \right] + \frac{q D \epsilon}{q-1} \left[ \langle n_{i-\sigma}^\sigma \rangle - \langle n_i^\sigma \rangle \right] + \sum_{\sigma' \ne \sigma } \left\langle n_i^{\sigma'} W_{\rm flip}(\sigma',\sigma) - n_i^\sigma W_{\rm flip}(\sigma,\sigma') \right\rangle.
\end{equation}

Now, we take the hydrodynamic limit for small lattice spacing $a\simeq 1/L$, corresponding to large system size limit $L\to\infty$. We define the average density of particles in the state $\sigma$ at the 2d position ${\bf x}$ as $\rho_\sigma({\bf x},t) = \langle n_i^\sigma(t) \rangle$ for which the coordinate ${\bf x}$ matches the lattice site $i$ at integer positions. In the appendices~\ref{appendixA} and \ref{appendixB}, we show that in this hydrodynamic limit the Master equation (\ref{MasterEq}) transforms into
\begin{equation}
\label{PDEhydro0}
\partial_t \rho_\sigma = D_\parallel \partial_\parallel^2 \rho_\sigma + D_\perp \partial_\perp^2 \rho_\sigma - v \partial_\parallel \rho_\sigma + \sum_{\sigma' \ne \sigma } I_{\sigma \sigma'},
\end{equation}
where
\begin{gather}
I_{\sigma \sigma'} = \left[\frac{q\beta J}{\rho}(\rho_\sigma+\rho_{\sigma'}) -1 - \frac{r}{\rho} - \alpha \frac{(\rho_\sigma-\rho_{\sigma'})^2}{\rho^2}\right](\rho_\sigma-\rho_{\sigma'}), \label{flipRMF} \\
D_{\parallel} = \frac{qD}{4}\left(1+\frac{\epsilon}{q-1}\right), \qquad D_{\perp} = \frac{qD}{4}\left(1-\frac{\epsilon}{q-1}\right), \qquad v=\frac{qD\epsilon}{q-1}.
\end{gather}
$D_{\parallel}$ and $D_{\perp}$ are the diffusion constants in the parallel direction ${\bf e_\parallel} = (\cos \phi_\sigma,\sin \phi_\sigma)$ and in the perpendicular direction ${\bf e_\perp} = (\sin \phi_\sigma, -\cos \phi_\sigma)$, respectively, with $\phi_\sigma=2 \pi (\sigma-1)/q$ the favored direction angle for a particle in state $\sigma$. $v$ is the self-propulsion velocity in the direction ${\bf e_\parallel}$. $\partial_\parallel = {\bf e_\parallel} \cdot \nabla$ and $\partial_\perp = {\bf e_\perp} \cdot \nabla$ are respectively the derivative in the parallel and perpendicular directions.

In the appendix \ref{appendixB}, we calculate the flipping term $I_{\sigma \sigma'}$ given in Eq.~(\ref{flipRMF}) where $\alpha= (q \beta J)^2(1-2\beta J/3)/2$ and $r= 6(q-1)^2\alpha_m \alpha/q^2$ depends only on the temperature $T=\beta^{-1}$. We take $ \gamma = \exp(-q\beta J/\rho_0)$ and we assume that the magnetization $m_\sigma$ in state $\sigma$, defined in the Eq.~(\ref{mag}), is small compared to the local density $\rho$. We keep only the ${\cal O}(m_\sigma^3)$ terms in a $m_\sigma \ll \rho$ expansion. Moreover, we assume that all magnetization are identically distributed Gaussian variables with variance $\alpha_m \rho$ proportional to the local mean density. This assumption is verified by MC simulations of the microscopic model shown in Fig.~\ref{figAPMalpha}. We see that $r$ simply rescales the densities $\rho_\sigma$ for which reason we can take $r=1$ without any loss of generality. In passing we note that the value $r=0$ (i.e. $\alpha_m=0$) corresponds to the conventional mean-field expression (without taking fluctuations into account) in which the magnetization are equal to their average values. The hydrodynamic equation (\ref{PDEhydro0}) can be rewritten as
\begin{equation}
\label{PDEhydro}
\partial_t \rho_\sigma = \nabla_{\bf x} \cdot {\frak D}_\sigma \nabla_{\bf x} \rho_\sigma - v {\bf e_\parallel} \cdot \nabla_{\bf x} \rho_\sigma + \sum_{\sigma' \ne \sigma } \left[\frac{q\beta J}{\rho}(\rho_\sigma+\rho_{\sigma'}) -1 - \frac{r}{\rho} - \alpha \frac{(\rho_\sigma-\rho_{\sigma'})^2}{\rho^2}\right](\rho_\sigma-\rho_{\sigma'}),
\end{equation}
with the diffusion tensor ${\frak D}_\sigma$ given by ${\rm diag}(D_\parallel,D_\perp)$ in the local frame of the state $\sigma$:
\begin{equation}
\label{defDsigma}
{\frak D}_\sigma = 
\begin{pmatrix}
\cos \phi_\sigma & -\sin \phi_\sigma \\
\sin \phi_\sigma & \cos \phi_\sigma
\end{pmatrix}
\begin{pmatrix}
D_\parallel & 0 \\
0 & D_\perp
\end{pmatrix}
\begin{pmatrix}
\cos \phi_\sigma & \sin \phi_\sigma \\
-\sin \phi_\sigma & \cos \phi_\sigma
\end{pmatrix}=
\frac{qD}{4} I_2 + \frac{qD\epsilon}{4(q-1)}
\begin{pmatrix}
\cos 2\phi_\sigma & \sin 2\phi_\sigma \\
\sin 2\phi_\sigma & - \cos 2\phi_\sigma
\end{pmatrix}
\end{equation}
where $\phi_\sigma = 2\pi (\sigma-1)/q$ and $I_2$ the identity matrix.

In passing we note that the $q\to\infty$ limit of this equation can be performed by rescaling the microscopic parameters as $\overline D = q D$, $\overline \epsilon = \epsilon/(q-1)$ and $\overline \gamma = q \gamma \exp(q \beta J/\rho_0)$ and introducing the average density of particles $\rho({\bf x},t; \phi)$ in the continuous state $\phi \in [0,2\pi]$ - which corresponds to $\phi_\sigma$ in the limit $q\to\infty$ - at the 2d position ${\bf x}$. In the appendix~\ref{appendixInfty} we obtain
\begin{gather}
\partial_t \rho(\phi) = \nabla_{\bf x} \cdot {\frak D}(\phi) \nabla_{\bf x} \rho(\phi) - v {e_\parallel}(\phi) \cdot \nabla_{\bf x} \rho(\phi) \nonumber\\ 
+ \overline \gamma \int_0^{2\pi} \frac{d \phi'}{2\pi} \left[\frac{2\pi\beta J}{\rho}(\rho(\phi)+\rho(\phi')) -1 - \frac{r}{\rho} - \overline \alpha \frac{(\rho(\phi)-\rho(\phi'))^2}{\rho^2}\right](\rho(\phi)-\rho(\phi')),
\end{gather}
where $\overline \alpha= (2 \pi \beta J)^2(1-2\beta J/3)/2$, $v= \overline D \overline \epsilon$, ${\bf e_\parallel}(\phi) = (\cos \phi, \sin \phi)$ and
\begin{equation}
{\frak D}(\phi) = \frac{\overline D}{4} I_2 +
\frac{\overline D \overline \epsilon}{4}
\begin{pmatrix}
\cos 2\phi & \sin 2\phi \\
\sin 2\phi & - \cos 2\phi
\end{pmatrix}.
\end{equation}

In the literature~\cite{toner,ST2015}, the hydrodynamic equation is usually formulated in terms of the total density $\rho({\bf x},t)$ and the polarization vector ${\bf P}({\bf x},t)$ defined as the average direction of the motion. In appendix~\ref{appendixPolarization} we show that the polarization vector can be expressed as
\begin{equation}
\label{polarization}
{\bf P} = \frac{q-1}{q}\sum_{\sigma=1}^q {\bf e_\sigma} \frac{m_\sigma}{\rho} = \sum_{\sigma=1}^q {\bf e_\sigma} \frac{\rho_\sigma}{\rho},
\end{equation}
with ${\bf e_\sigma} = (\cos \phi_\sigma,\sin \phi_\sigma)$ corresponding to ${\bf e_\parallel}$ for the state $\sigma$. The polarization vector used for active models is then linked to the magnetization $m_\sigma$ defined naturally for the Potts model. Taking the sum over all states in Eq.~(\ref{PDEhydro}), we obtain for the density $\rho$
\begin{equation}
\partial_t \rho = \sum_{\sigma=1}^q \nabla_{\bf x} \cdot {\frak D}_\sigma \nabla_{\bf x} \rho_\sigma - v \sum_{\sigma=1}^q {\bf e_\sigma} \cdot \nabla_{\bf x} \rho_\sigma,
\end{equation}
where the flipping terms canceled due to the anti symmetry of $I_{\sigma \sigma'}$. Using the Eq.~(\ref{polarization}) and the expression of ${\frak D}_\sigma$ given by Eq.~(\ref{defDsigma}), we get
\begin{equation}
\partial_t \rho = \frac{qD}{4} \nabla_{\bf x}^2 \rho - v \nabla_{\bf x} \cdot \left( \rho {\bf P} \right) + \frac{v}{4} \left[ (\partial_x^2 - \partial_y^2) \sum_{\sigma=1}^q \cos(2\phi_\sigma) \rho_\sigma + 2 \partial_x \partial_y \sum_{\sigma=1}^q \sin(2\phi_\sigma) \rho_\sigma \right].
\end{equation}
The last term of this equation is not present in the Toner and Tu equations~\cite{toner}, and cannot by expressed with the polarization vector. Hence the APM appears not to fall into the universality classes described by the Toner-Tu model.

\subsection{Numerical solutions}

\begin{figure}[t]
\begin{minipage}[t]{.49\linewidth}
\begin{center}
\includegraphics[width=8cm]{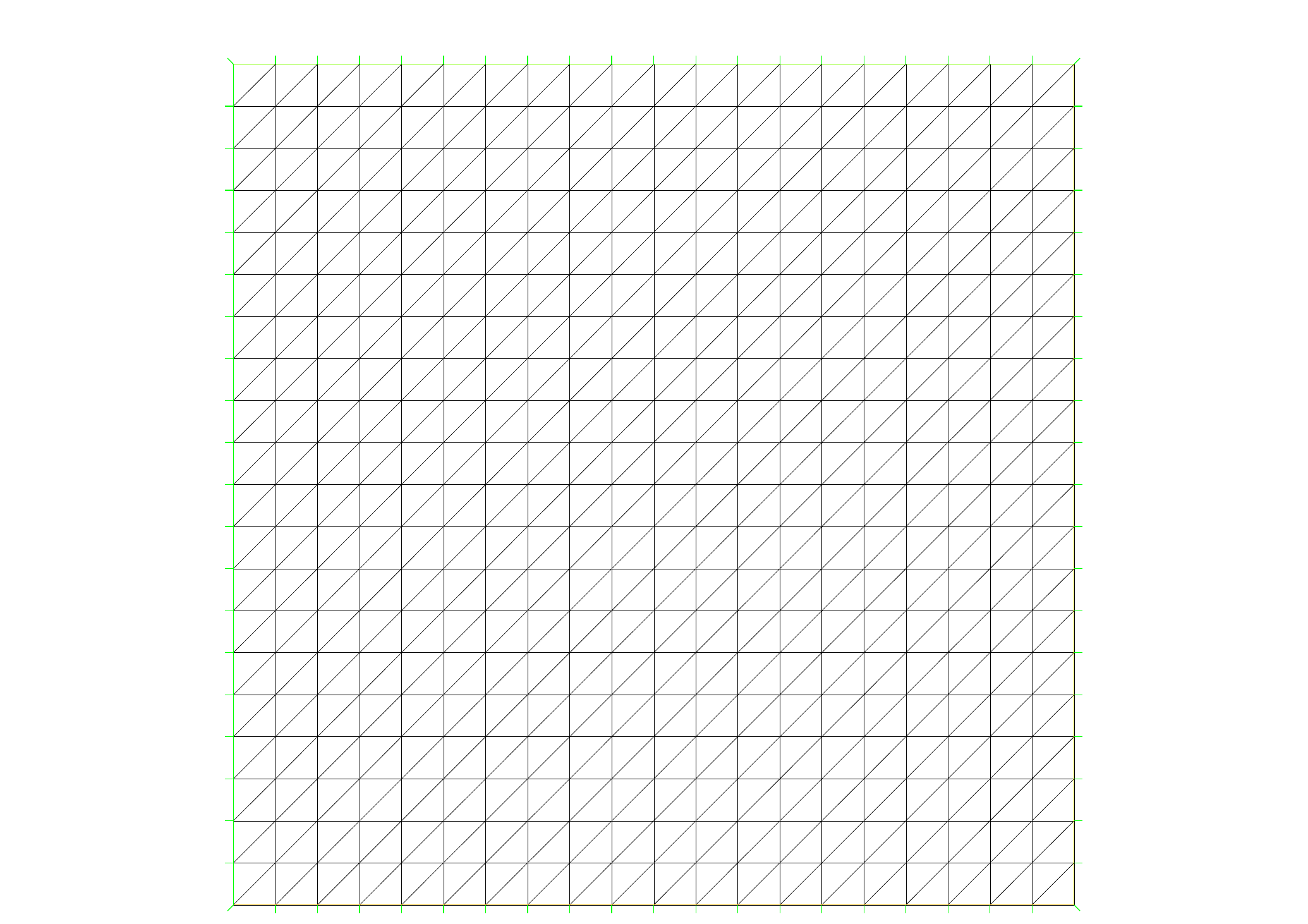} 
\end{center}
\end{minipage}
\begin{minipage}[t]{.49\linewidth}
\begin{center}
\includegraphics[width=8cm]{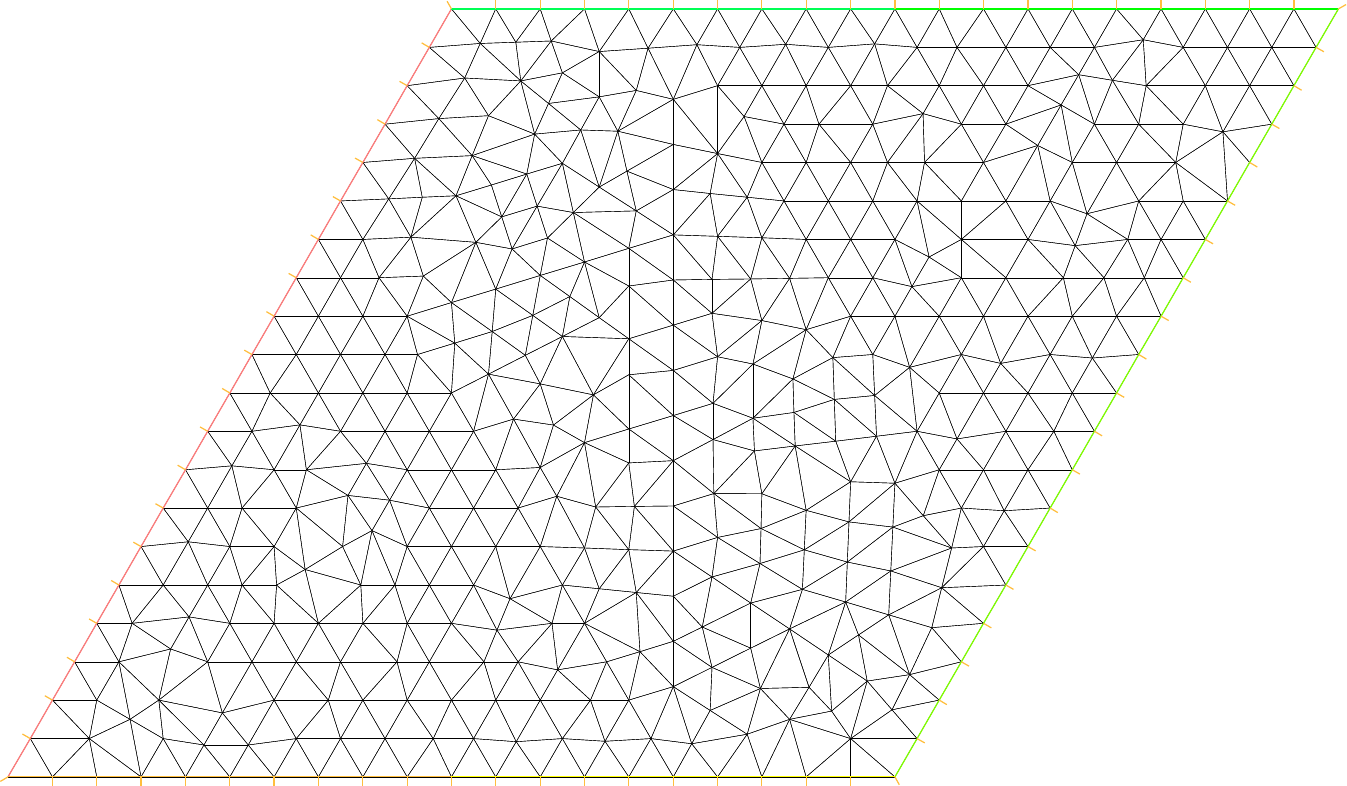}
\end{center}
\end{minipage}
\caption{Triangular mesh-grids taken to integrate the hydrodynamic equation (Eq.~(\ref{PDEhydro})) for $q=4$ (left) and $q=6$ (right) are represented here with ${\cal N}=20$ vertices on each border.\label{fig_APM_mesh}}
\end{figure}

We solve Eq.~(\ref{PDEhydro}) numerically and set $D=1$, $J=1$ and $r=1$ (without any loss of generality) defining the scaling of time, temperature and density. We use FreeFEM++~\cite{ff}, a software package based on the finite element method~\cite{fem}. The equations are integrated for a discrete time with time increments $\Delta t$ such that $t_n = n \Delta t$ is the time at the $n^{\rm th}$ step. We define $\rho_\sigma^{(n)}({\bf x})= \rho_\sigma({\bf x},t_n)$ as the density at the discrete time $t_n$. As initial condition $\rho_\sigma^{(0)}$ we take a high density bubble or stripe (horizontal or vertical) with non straight boundaries in a low density phase. The low density phase is a gas phase ($m_\sigma=0$) and the high density phase a polar liquid phase in the state $\sigma=1$ ($m_1>0$). Eq.~(\ref{PDEhydro}) can be rewritten as
\begin{equation}
\rho_\sigma^{(n+1)}-\rho_\sigma^{(n)} = \Delta t \left[ \nabla_{\bf x} \cdot {\frak D}_\sigma \nabla_{\bf x} \rho_\sigma^{(n+1)} - v {\bf e_\parallel} \cdot \nabla_{\bf x} \rho_\sigma^{(n+1)} + \sum_{\sigma' \ne \sigma } K_{\sigma \sigma'}^{(n)}(\rho_\sigma^{(n+1)}-\rho_{\sigma'}^{(n+1)}) \right],
\end{equation}
with $\rho_\sigma^{(n)}({\bf x})$ the known particle density at time $t_n$, $\rho_\sigma^{(n+1)}({\bf x})$ the unknown particle density at time $t_{n+1}$ and $K_{\sigma \sigma'}^{(n)}({\bf x})$ the (symmetric) quantity defined by
\begin{equation}
K_{\sigma \sigma'} = \frac{q\beta J}{\rho}(\rho_\sigma+\rho_{\sigma'}) -1 - \frac{r}{\rho} - \alpha \frac{(\rho_\sigma-\rho_{\sigma'})^2}{\rho^2}
\end{equation}
evaluated at time $t_n$. The final time is denoted $t_{\rm max}$. From the Lax-Milgram theorem \cite{LaxMilgram}, these linear equations have unique solutions $\rho_\sigma^{(n+1)}({\bf x})$ for the $(n+1)^{\rm th}$ step. The weak formulation of these equations is the integral equation:
\begin{gather}
\int_\Omega d{\bf x} \ \left[ \sum_\sigma w_\sigma \rho_\sigma^{(n+1)} + \Delta t \sum_\sigma \nabla_{\bf x} w_\sigma \cdot {\frak D}_\sigma \nabla_{\bf x} \rho_\sigma^{(n+1)} \right. \nonumber\\
\left.+ v \Delta t \sum_\sigma w_\sigma {\bf e_\parallel} \cdot \nabla_{\bf x} \rho_\sigma^{(n+1)} - \Delta t \sum_{\sigma' > \sigma } (w_\sigma - w_{\sigma'}) K_{\sigma \sigma'}^{(n)} \left(\rho_\sigma^{(n+1)}-\rho_{\sigma'}^{(n+1)} \right)\right] = \int_\Omega d{\bf x} \ \sum_\sigma w_\sigma \rho_\sigma^{(n)},
\end{gather}
where $w_\sigma({\bf x})$ are test functions. This equation can be written in the form $b(w_\sigma,\rho_\sigma) = l(w_\sigma)$, where $b(w_\sigma,\rho_\sigma)$ is a bilinear function and $l(w_\sigma)$ is a linear function of the space of integrable functions ($L^1$). To solve this integral equation, space is discretized into a triangular mesh-grid with ${\cal N}$ vertices on the boundaries. A representation of these grids used for $q=4$ and $q=6$ is shown in Fig.~\ref{fig_APM_mesh} for ${\cal N}=20$ vertices. Note that the precision of the numerical solution is increased for a narrow grid (${\cal N}\gg 1$) and small time increments ($\Delta t \ll 1$).

The functions are then calculated at the nodes of the mesh-grid and interpolated linearly over the complete space. The number of nodes is of order ${\cal O}({\cal N}^2)$, which corresponds to the number of unknowns in the numerical problem (for each state $\sigma$). The interpolation is made with the help of the Lagrange polynomials $e_i({\bf x})$ forming a base on the discretized space, and defined by its value on the nodes: $1$ at the $i^{\rm th}$ node and $0$ at other nodes. Then the density function is $\rho_\sigma({\bf x}) = \sum_i \rho_{\sigma,i} \ e_i({\bf x})$ for a value $\rho_{\sigma,i}$ at the $i^{\rm th}$ node. Using the linearity of the integral equation and replacing the test functions by Lagrange polynomials we get $\sum_j \rho_{\sigma,j} b(e_i,e_j) = l(e_i)$ for all $i$. This can be rewritten using the matrix formulation as $M R = V$ where $M$ is a matrix with elements $M_{ij} = b(e_i,e_j)$ and $R$, $V$ are vectors such that $R_j=\rho_{\sigma,j}$ and $V_i=l(e_i)$. The solution is then given by $R = M^{-1} V$. The computational time has a complexity proportional to the number of nodes (order ${\cal N}^2$) and it takes approximately $20$ hours for ${\cal N}=100$ and $t_{\rm max}/\Delta t = 10^4$ time steps, on a 4 GHz processor.

\begin{figure}[t]
\begin{center}
\includegraphics[width=16cm]{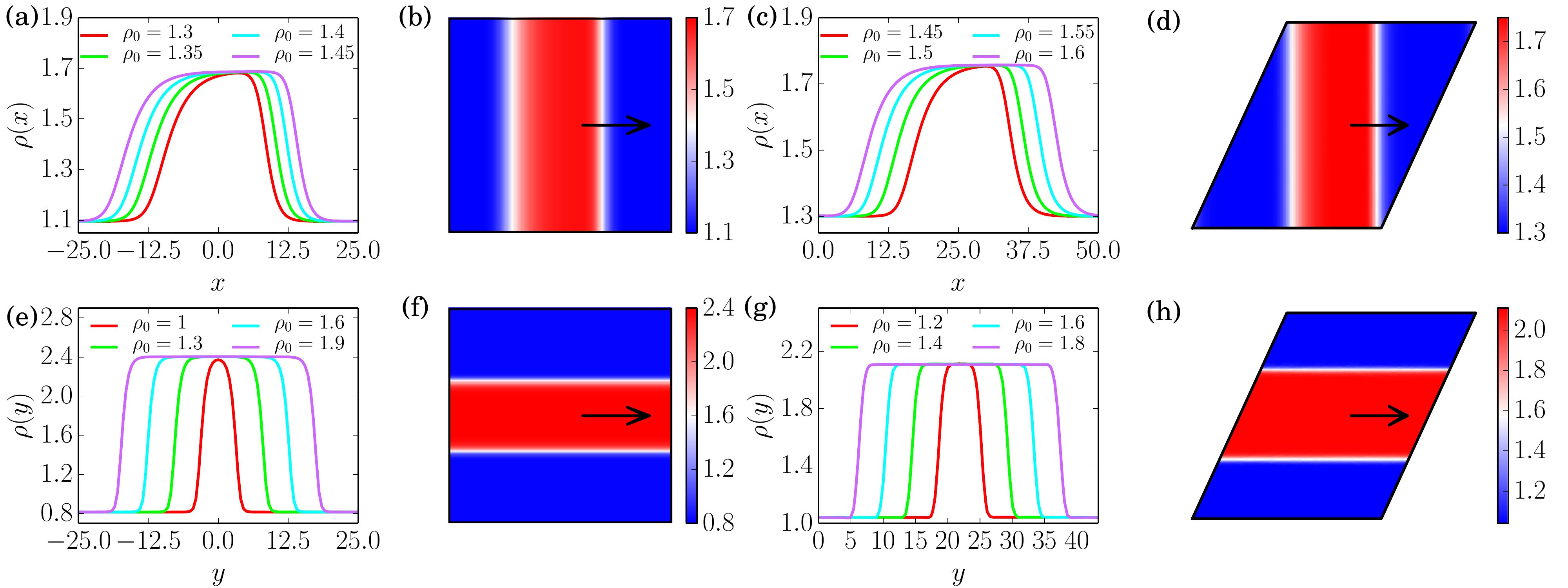} 
\caption{(color online) Stationary profiles (a,c,e,g) and snapshots of the bands (b,d) and lanes (f,h) (red and blue denote the liquid and gas phases, respectively) of the APM. (a-b) Band motion for $q=4$, $\beta=0.75$ and $\epsilon=0.5$ where $\rho_{\rm gas} = 1.10$ and $\rho_{\rm liq} = 1.69$, and (c-d) for $q=6$, $\beta=0.65$ and $\epsilon=0.5$ where $\rho_{\rm gas} = 1.30$ and $\rho_{\rm liq} = 1.75$. (e-f) Lane formation for $q=4$, $\beta=0.75$ and $\epsilon=2.5$ where $\rho_{\rm gas} = 0.815$ and $\rho_{\rm liq} = 2.40$, and (g-h) for $q=6$, $\beta=0.65$ and $\epsilon=4.5$ where $\rho_{\rm gas} = 1.04$ and $\rho_{\rm liq} = 2.11$. The reorientation transition occurs for $q=4$ at $\epsilon=2.0$ between (b) and (f) plotted for $\rho_0=1.33$, and for $q=6$ at $\epsilon=3.8$ between (d) and (h) plotted for $\rho_0=1.5$. The linear system size is $L=50$ and the numerical parameters are $\Delta t=0.2$ and ${\cal N} = 100$. \label{fig_APM_bands}}
\end{center}
\end{figure}

In the Fig.~\ref{fig_APM_bands}(a-b,e-f), we show the examples of the stationary phase-separated profiles obtained for the 4-state APM with $\beta=0.75$, $L=50$ and two values of $\epsilon$: $\epsilon=0.5$ shows a transverse band and $\epsilon=2.5$ a longitudinal lane. Similarly, in Fig.~\ref{fig_APM_bands}(c-d,g-h), we show the corresponding data for the 6-state APM with $\beta=0.65$, $L=50$ and two values of $\epsilon$: $\epsilon=0.5$ shows a transverse band and $\epsilon=4.5$ a longitudinal lane. For all these solutions, the numerical parameters are $\Delta t = 0.2$ and ${\cal N}=100$, and the stationary profiles are plotted at the time $t_{\rm max} = {\cal O}(10^3)$, chosen sufficiently large to obtain the stationary profile depending on the initial conditions and physical parameters. For small values of $\epsilon$ we observe in Fig.~\ref{fig_APM_bands}(a--d) a band motion $\rho(x-ct)$ with a velocity $c$ larger than the self-propulsion velocity $v$, presenting an asymmetric profile along $x$. For large values of $\epsilon$ we observe in Fig.~\ref{fig_APM_bands}(e--h) a lane formation $\rho(y)$ moving with a velocity $v$, appearing as a symmetric immobile profile. This behaviour indicates the presence of a reorientation transition which takes place at $\epsilon = \epsilon_*$. For these parameters, we note that $\epsilon_* \simeq 2.0$ for $q=4$ and $\epsilon_* \simeq 3.8$ for $q=6$.

The reorientation transition can occur on the basis of the characteristic times of each microscopic process: the ballistic transport $\tau_{\rm ballistic} \sim 1/v$, the longitudinal diffusion $\tau_{\rm diff}^\parallel \sim 1/D_\parallel$, the transverse diffusion $\tau_{\rm diff}^\perp \sim 1/D_\perp$ and the spin-flip from an unfavoured to a favoured spin orientation $\tau_{\rm flip} \sim \exp(-q\beta J)$. In the temperature range $T\in[1,5]$, the characteristic flip time is around $1/2^q$, which is smaller than the longitudinal diffusion time $\tau_{\rm diff}^\parallel$. For small values of $\epsilon$, one has $\tau_{\rm diff}^\perp \sim \tau_{\rm diff}^\parallel \ll \tau_{\rm ballistic}$, implying that the time scales for transverse and longitudinal diffusion are both of order one. The situation is then similar to the AIM analyzed in \cite{ST2015} in the limit $\epsilon \rightarrow 0$, allowing only transverse bands. On the other hand, for large values of $\epsilon$ (i.e. close to $q-1$), one has $\tau_{\rm diff}^\perp \gg \tau_{\rm ballistic}$, implying that the longitudinal ballistic transport is faster than the transverse diffusion allowing longitudinal lanes. In other words, for large velocities, a weakly perturbed longitudinal lane will be stable, in which case any transverse perturbation will vanish due to the fast longitudinal ballistic transport. In contrast, for small velocities, a weakly perturbed transverse band will be stable, in which case the longitudinal perturbation will vanish due to the faster diffusive process. A rigorous quantitative explanation of this reorientation transition will be presented in section \ref{section4}, based on a linear stability analysis of the homogeneous solutions.

\begin{figure}[t]
\begin{center}
\includegraphics[width=16cm]{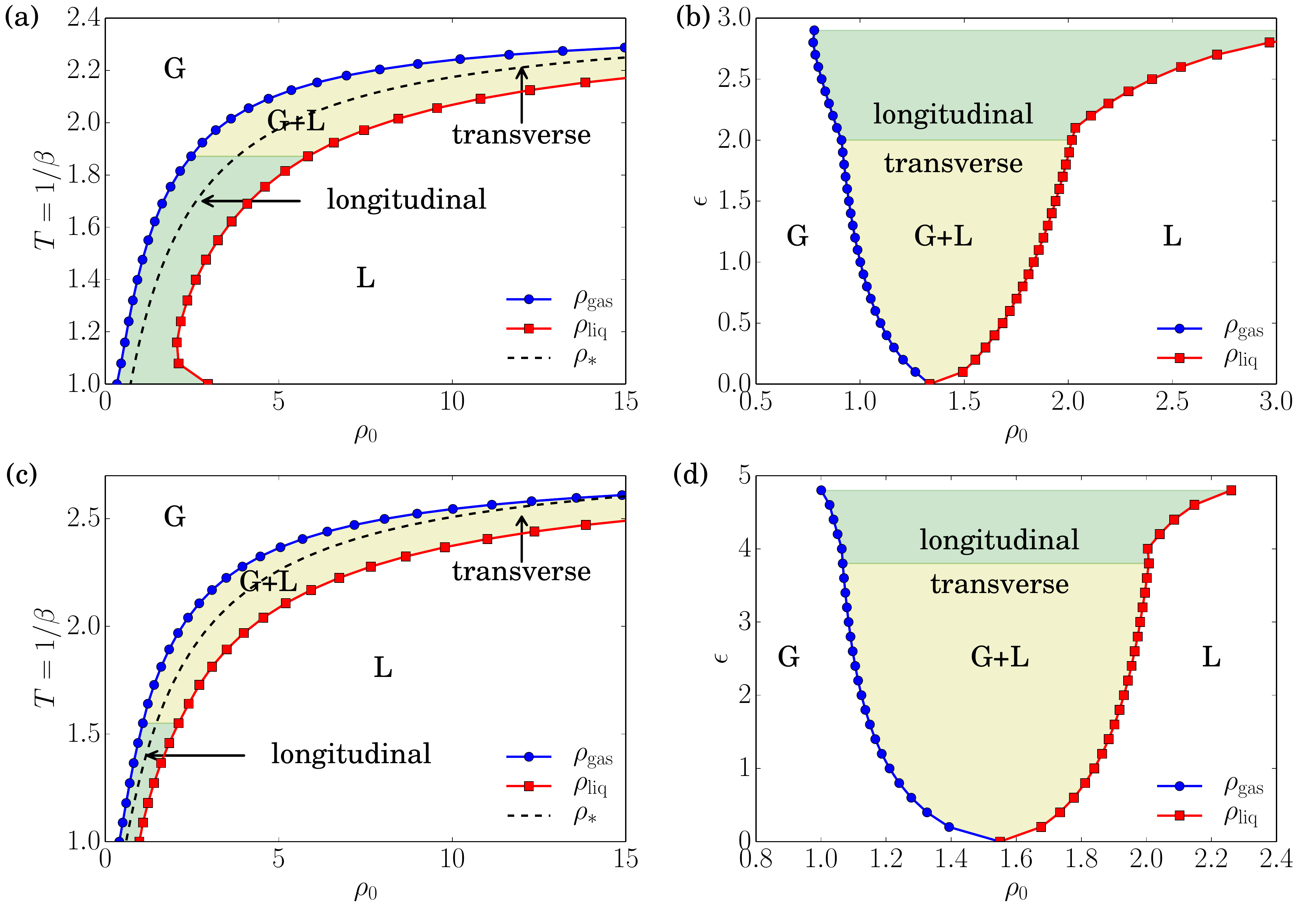}
\end{center}
\caption{(color online) Phase diagrams of the q-state APM: (a) Temperature-density phase diagram for $q=4$ and $\epsilon=2.5$ and (c) for $q=6$ and $\epsilon=4$. (b) Velocity-density phase diagram for $q=4$ and $\beta=0.75$ and (d) for $q=6$ and $\beta=0.65$. The linear system size is $L=100$ and the numerical parameters are $\Delta t=0.1$ and ${\cal N}=200$. The stationary state is reached for the final time $t_{\rm max} = 1000$. In all phases diagrams the regions where transverse band motion and longitudinal lane formation are stable within the phase separated domain (G+L) are shown. \label{fig_APM_diagrams}}
\end{figure}

In Fig.~\ref{fig_APM_diagrams}(a-b) we present the temperature-density (for $\epsilon=2.5$) and the velocity-density (for $\beta=0.75$) diagrams for the 4-state APM. The binodals $\rho_{\rm gas}$ and $\rho_{\rm liq}$ separate the gas (G) and liquid (L) domains from the phase-separated domain (G+L) while the line $\rho_*$ represents the ordered-disordered transition at $\epsilon=0$. In the (G) and (L) domains, the disordered and ordered homogeneous solutions are stable while in the (G+L) domain, inhomogeneous profiles can be observed. The values of these binodals are obtained from the stationary phase-separated profiles, representing the lower and the higher values of the density profile. Two different inhomogeneous profiles can be seen for the APM: a transverse band of polar liquid at small $\epsilon$ and large $T$ values and a longitudinal lane of polar liquid at large $\epsilon$ and small $T$ values. For fixed $\epsilon=2.5$ the reorientation transition occurs at $\beta=0.53$ (independent of the density $\rho_0$, c.f. Fig.~\ref{fig_APM_diagrams}(a)) and for fixed $\beta=0.75$ at $\epsilon=2.0$ (independent of $\rho_0$, c.f. Fig.~\ref{fig_APM_diagrams}(b)). Note that the liquid phase does not appear for $T>T_c$, leading to a liquid-gas phase diagram with a critical point located at $T_c \simeq 2.4$ and $\rho_0=+\infty$ as already described in Ref.~\cite{ST2015} for the AIM. Fig.~\ref{fig_APM_diagrams}(c-d) shows the temperature-density (for $\epsilon=4$) and the velocity-density (for $\beta=0.65$) diagrams for the 6-state APM. We obtain similar liquid-gas phase diagrams as for $q=4$, with a critical temperature $T_c \simeq 2.8$. The reorientation transition takes place for $\beta=0.65$ at $\epsilon=4.0$, c.f. Fig.~\ref{fig_APM_diagrams}(c) and Fig.~\ref{fig_APM_diagrams}(d).


\section{Homogeneous solutions and linear stability}
\label{section4}

The homogeneous solutions, $\rho_\sigma({\bf x})=\rho_\sigma={\rm const.}$, must satisfy the relation $\rho = \rho_0$ in order to fulfill $\int d{\bf x} \ \rho({\bf x},t)/L^2 = \rho_0$. A trivial homogeneous solution for Eq.~(\ref{PDEhydro}) is given by $\rho_{\sigma'} = \rho_{\sigma}$ for all pairs $({\sigma},{\sigma'})$ implying $\rho_{\sigma} = \rho_0/q$. The magnetization corresponding to this solution are $m_{\sigma} = 0$, corresponding to the disordered homogeneous solution. Next we examine if an ordered homogeneous solution exists. We consider a broken symmetry phase favoring right moving particles (spin-state $\sigma = 1$) corresponding to a positive magnetization $m_{\sigma=1} = m_0$ and a density $\rho_{\sigma=1}$ larger than the density of the other states. All other states have the same magnetization: $-m_0/(q-1)$ such that the total magnetization is zero. Then we have $\rho_\sigma = (\rho_0-m_0)/q + m_0$ and $\rho_{\sigma'} = (\rho_0-m_0)/q$, implying that $\rho_{\sigma}-\rho_{\sigma'} = m_0$ and $\rho_{\sigma}+\rho_{\sigma'} = 2(\rho_0-m_0)/q + m_0$. From the flipping term $I_{\sigma,\sigma'}$ given by Eq.~(\ref{flipRMF}), the magnetization $m_0$ must satisfy the equation
\begin{equation}
\left[ 2\beta J - 1 - \frac{r}{\rho_0} + (q-2)\beta J \frac{m_0}{\rho_0} - \alpha \frac{m_0^2}{\rho_0^2} \right] m_0 = 0.
\end{equation}
This equation has three different solutions: (i) $m_0=0$ corresponding to the trivial disordered solution and (ii) two ordered solutions with
\begin{equation}
\label{eqMag}
M \equiv \frac{m_0}{\rho_0} = \frac{(q-2)\beta J}{2\alpha} \left\{ 1 \pm \sqrt{1+ \frac{4 \mu_0 \alpha}{(q-2)^2(\beta J)^2}} \right\},
\end{equation}
where $\mu_0 = 2\beta J - 1 -r/\rho_0$ and $\alpha= (q \beta J)^2(1-2\beta J/3)/2$. These ordered homogeneous solutions exists only when $4 \mu_0 \alpha+(q-2)^2 (\beta J)^2>0$ i.e. when
\begin{equation}
\label{defrhostar}
\rho_0 > \frac{2q^2(1-2\beta J/3)r}{(q-2)^2+2q^2 (2\beta J-1)(1-2\beta J/3)} \equiv \rho_*
\end{equation}
which defines the density $\rho_*$ corresponding to the minimal value of $\rho_0$ for which an ordered homogeneous solution exists. Additionally, the temperature must satisfy the relations $(q-2)^2+2q^2 (2\beta J-1)(1-2\beta J/3) > 0$ and $1-2\beta J/3>0$ giving
\begin{equation}
T_c(q)^{-1} \equiv 1- \frac{\sqrt{6q^2 -6q^3 +5q^4/2}}{2q^2} < \beta J < \frac{3}{2}.
\end{equation}
This inequality implies that the ordered liquid phase does not exist for temperatures larger than $T_c(q)$, corresponding to the critical temperature of the liquid-gas transition. $T_c(q)$ is a strictly increasing function of $q$ with special values: $T_c(2) = 2$, $T_c(4) = (1-\sqrt{22}/8)^{-1} \simeq 2.417$, $T_c(6) = (1-\sqrt{5/12})^{-1} \simeq 2.820$ and $T_c(+\infty) = (1-\sqrt{5/8})^{-1} \simeq 4.770$.

The magnetization given by Eq.~(\ref{eqMag}) can be rewritten by using 
\begin{equation}
1+ \frac{4 \mu_0 \alpha}{(q-2)^2(\beta J)^2} = \frac{\alpha r}{(q-2)^2(\beta J)^2} \frac{\rho_0-\rho_*}{\rho_* \rho_0}.
\end{equation}
derived from Eq.~(\ref{defrhostar}) defining $\rho_*$. Then, we obtain for the magnetization
\begin{equation}
\label{eqMag2}
M = \frac{(q-2)\beta J}{2\alpha} \pm \sqrt{\frac{r}{\alpha \rho_*}} \sqrt{\frac{\rho_0-\rho_*}{\rho_0}} \equiv M_0 \pm M_1 \delta,
\end{equation}
where $M_0=(q-2)\beta J / \alpha$ and $M_1 = \sqrt{r/\alpha \rho_*}$ are temperature dependent constants and $\delta = \sqrt{(\rho_0-\rho_*)/\rho_0}$ is a variable with values between $0$ and $1$. When $\rho_0 = \rho_*$ (i.e. $\delta=0$), the magnetization is equal to $M_0$ and increases (decreases) with $\rho_0$ through the maximal (minimal) value given by $M_0 + M_1$ ($M_0-M_1$). Depending on the temperature, the maximal value can be larger than $1$ ($M_0+M_1>1$) and the minimal value is always negative ($M_0-M_1<0$).

\begin{figure}[t]
\begin{center}
\includegraphics[width=16cm]{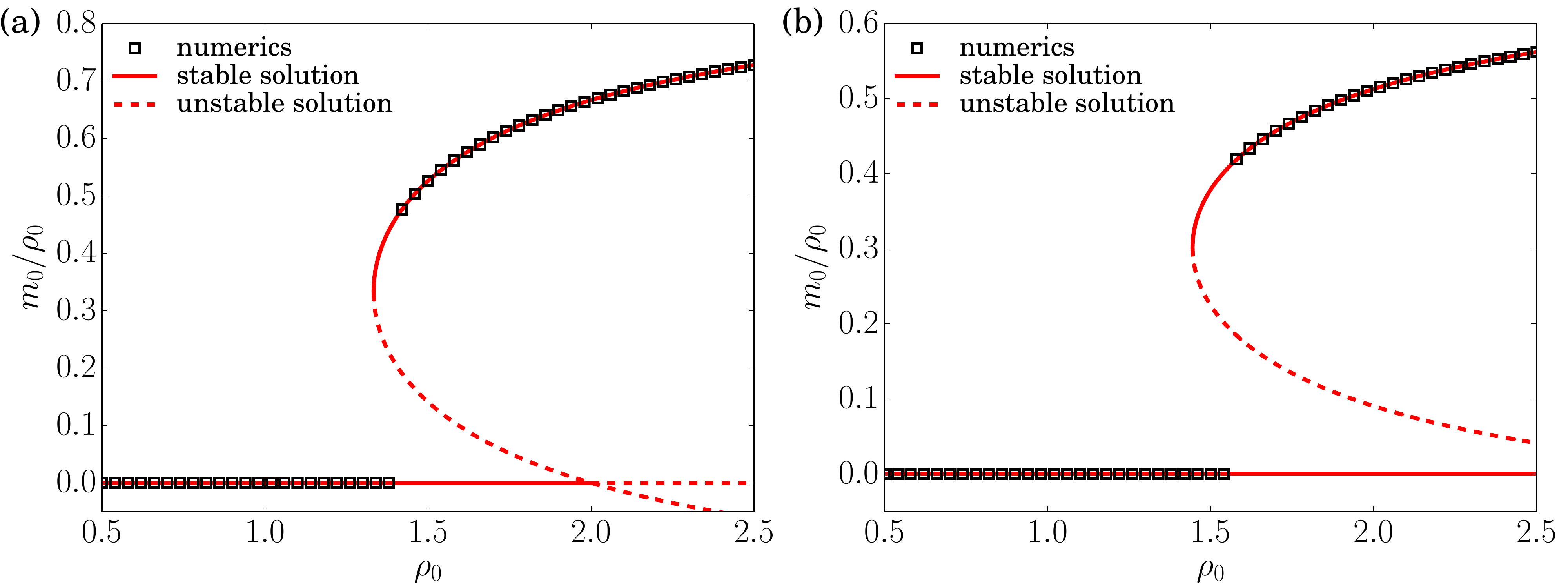} 
\end{center}
\caption{Normalized magnetization $m_0/\rho_0$ of the homogeneous solutions for $q=4$ and $\beta=0.75$ in (a), and $q=6$ and $\beta=0.65$ in (b) as a function of the average density $\rho_0$. We show the analytical solutions given by Eq.~(\ref{eqMag}) and the stability of these solutions analyzed in sections \ref{section4a} and \ref{section4b} for the disordered and ordered solutions, respectively. The square symbols represent the stationary state of the numerical solution for $L=100$ and the numerical parameters $\Delta t=0.2$ and ${\cal N}=200$, starting from a phase separated profile at $\epsilon=0$ for which the stationary state is homogeneous for all density values. \label{fig_APM_magnetisation}}
\end{figure}

In Fig.~\ref{fig_APM_magnetisation}, we represent the normalized magnetization $m_0/\rho_0$ of homogeneous solutions for $q=4$ and $q=6$ as a function of the average density $\rho_0$ which follows the Eq.~(\ref{eqMag}). The transition between the disordered phase ($m_0=0$) and the ordered phases ($m_0 \ne 0$) is a transcritical type bifurcation instead of a pitchfork bifurcation for the active Ising model ($q=2$) and takes place at the density $\rho_*$. We represent also the stationary state obtained numerically starting from a phase-separated profile at $\epsilon=0$, determining the most stable homogeneous phase. We remark that the numerical ordered-disordered transition happens for a density larger than $\rho_*$, due to the transcritical property of the transition.

Now, we look at the stability of the three different homogeneous solutions. We add a small perturbation $\delta \rho_\sigma({\bf x},t)$ to the homogeneous solution. We expand the hydrodynamic equations to first order in this perturbation $\delta\rho_\sigma$ and analyze the time-evolution of the perturbation of the disordered homogeneous solution in section \ref{section4a} and of the ordered solutions in section \ref{section4b}. If the perturbation vanishes at large times, then the homogeneous solution is stable.

\subsection{Linear stability of disordered homogeneous solution}
\label{section4a}

Adding a small perturbation to the disordered homogeneous solution, the particle density is given by $\rho_\sigma({\bf x}, t) = \rho_0/q + \delta \rho_\sigma({\bf x}, t)$ for all states $\sigma$. From Eq.~(\ref{flipRMF}), the flipping term is up to first order in $\delta \rho_\sigma$:
\begin{equation}
I_{\sigma \sigma'} = \left[ 2\beta J -1 - \frac{r}{\rho_0} \right](\delta\rho_{\sigma}-\delta\rho_{\sigma'}) = \mu_0 (\delta\rho_{\sigma}-\delta\rho_{\sigma'}),
\end{equation}
where $\mu_0 = 2\beta J - 1 -r/\rho_0$ as defined previously. Taking the 2d Fourier transform of $\delta \rho_\sigma({\bf x},t)$ such that
\begin{equation}
\label{FourierTrans}
\delta \rho_\sigma({\bf x},t) = \int \frac{d{\bf k}}{(2\pi)^2} \exp(i {\bf k} \cdot {\bf x}) \widehat{\delta\rho_\sigma}({\bf k},t),
\end{equation}
the Eq.~(\ref{PDEhydro}) gives the evolution of $\widehat{\delta\rho_\sigma}({\bf k},t)$ following the equation
\begin{equation}
\partial_t \widehat{\delta\rho_{\sigma}} = \left(-{\bf k} \cdot {\frak D}_\sigma {\bf k} - i v {\bf k} \cdot {\bf e_\parallel} \right) \widehat{\delta\rho_{\sigma}} + \mu_0 \sum_{{\sigma'} \ne {\sigma} } \left(\widehat{\delta\rho_{\sigma}} - \widehat{\delta\rho_{\sigma'}} \right).
\end{equation}
Denoting $R=\begin{pmatrix}
\widehat{\delta\rho_1} & \dots & \widehat{\delta\rho_q}
\end{pmatrix}^\intercal$, this last equation rewrites as $\partial_t R = M_{\rm gas} R$ with the matrix $M_{\rm gas}$ defined by its components
\begin{equation}
\label{defMgas}
\left[M_{\rm gas}\right]_{\sigma \sigma'} = \left(-{\bf k} \cdot {\frak D}_\sigma {\bf k} - i v {\bf k} \cdot {\bf e_\parallel} + q \mu_0\right) \delta_{\sigma \sigma'} - \mu_0,
\end{equation}
with $\delta_{\sigma \sigma'}$ the Kronecker delta. Since the matrix $M_{\rm gas}$ is symmetric, it is diagonalizable. We define $\lambda_{\rm gas}^i$ the eigenvalues of this matrix. Then $M_{\rm gas} = P^{-1} \Delta P$, where $\Delta_{ij} = \lambda_{\rm gas}^i \delta_{ij}$ is a diagonal matrix and $P$ is the matrix of the eigenvectors. The evolution is then given by $R(t) = \exp(Mt) R(0)$ where the exponential of the matrix is expressed in terms of the eigenvalues as $R_i(t) = P^{-1}_{ij} \exp(\lambda_{\rm gas}^j t) P_{jk} R_k(0)$. In the eigenspace the evolution of the vector $P R(t)$ is exponential. Then, the perturbation vanishes only when the real part of {\itshape all} eigenvalues is negative.

In the appendices \ref{appendixC1} and \ref{appendixD1}, we show that the disordered solution is stable when $\mu_0<0$, for $q=4$ and $q=6$ respectively. This inequality is equivalent to $r/\rho_0>2\beta J -1$. For $\beta J<1/2$, this relation is always verified and for $\beta J>1/2$, this relation implies that
\begin{equation}
\rho_0 < \frac{r}{2\beta J-1} \equiv \varphi_{\rm gas}
\end{equation}
which defines the gas spinodal $\varphi_{\rm gas}$, the maximal density for which the disordered homogeneous solution is stable. Note that $\varphi_{\rm gas} = + \infty$ for $\beta J<1/2$. We can then write the relation with $\rho_*$ (for $q>2$):
\begin{equation}
\frac{\varphi_{\rm gas}-\rho_*}{r} = \left[ 2 \beta J -1 + \frac{2q^2}{(q-2)^2}(2\beta J -1)^2 (1-2\beta J/3) \right]^{-1} >0.
\end{equation}
Then $\varphi_{\rm gas}$ is larger that $\rho_*$ for all temperatures, then the disordered solution stays stable when the ordered solution appears for $q>2$. The ordered and disordered solutions may be both stable at a given density. In Fig.~\ref{fig_APM_magnetisation}, the stability of the disordered solution is represented for $q=4$ and $q=6$ and the transcriticality property of the bifurcation is compatible with the relation $\varphi_{\rm gas}> \rho_*$. When $q=2$, we recall that $\varphi_{\rm gas} = \rho_*$ \cite{ST2015}.

\subsection{Linear stability of ordered homogeneous solutions}
\label{section4b}

Adding a small perturbation to the ordered homogeneous solution, the density of particles in state $\sigma=1$ is given by $\rho_\sigma({\bf x},t) = (\rho_0 - m_0)/q + m_0 + \delta \rho_\sigma ({\bf x},t)$ while the density of particles in the other states is $\rho_{\sigma'}({\bf x},t) = (\rho_0- m_0)/q + \delta \rho_{\sigma'} ({\bf x},t)$. We denote in the following $\delta \rho ({\bf x},t) = \sum\limits_{\sigma=1}^q \delta \rho_{\sigma} ({\bf x},t)$, the perturbation to the total density. The flipping term defined by the Eq.~(\ref{flipRMF}) has two different components: one for the flip from spin-state $1$ to any other state ${\sigma}$, denoted by $I^{(1)}_{1 \sigma}$, and another for the flip between two states $\sigma$ and $\sigma'$ different from spin-state $1$, denoted by $I^{(2)}_{\sigma \sigma'}$. Only the first term on the r.h.s. of Eq.~(\ref{flipRMF}) gives a linear contribution of the perturbation $\delta \rho_\sigma$ to $I^{(1)}_{1 \sigma}$. It follows that
\begin{equation}
I^{(1)}_{1 \sigma} = \left[ q \beta J (\delta \rho_1 + \delta \rho_{\sigma}) - 2 \beta J \left(1 - \frac{q-2}{2}\frac{m_0}{\rho_0} \right) \delta \rho + \frac{r}{\rho_0} \delta \rho -2 \alpha \frac{m_0}{\rho_0} (\delta \rho_1 - \delta \rho_{\sigma}) + 2 \alpha \frac{m_0^2}{\rho_0^2} \delta \rho\right] \frac{m_0}{\rho_0}.
\end{equation}
Denoting $M=m_0 / \rho_0$ and using the relation $\mu_0 + (q-2)\beta J M - \alpha M^2$, we get
\begin{equation}
\label{expI1}
I^{(1)}_{1 \sigma} = M (q\beta J - 2 \alpha M) \delta \rho_1 + M (q\beta J + 2 \alpha M) \delta \rho_{\sigma} + (\alpha M^2-1) \delta \rho.
\end{equation}

The contribution to $I^{(2)}_{\sigma \sigma'}$ linear in the perturbation $\delta \rho_\sigma$ derives from the second term on the r.h.s. of Eq.~(\ref{flipRMF}). Then using the definition of $M$, it follows that
\begin{equation}
\label{expI2}
I^{(2)}_{\sigma \sigma'} = \left[ 2 \beta J - 1 - \frac{r}{\rho_0} -2 \beta J \frac{m_0}{\rho_0} \right] (\delta \rho_{\sigma} - \delta \rho_{\sigma'}) = M (\alpha M - q \beta J) (\delta \rho_{\sigma} - \delta \rho_{\sigma'}).
\end{equation}

Taking the 2d Fourier transform of $\delta \rho_\sigma({\bf x},t)$ defined by Eq.~(\ref{FourierTrans}) and using the Eqs.~(\ref{expI1}) and~(\ref{expI2}), the evolution of $\widehat{\delta \rho_\sigma}$ given by the Eq.~(\ref{PDEhydro}) becomes for the $\sigma=1$ state
\begin{equation}
\partial_t \widehat{\delta \rho_1} = \left( - {\bf k} \cdot {\frak D}_1 {\bf k} - i v {\bf k} \cdot {\bf e_\parallel} + (q-1) \mu \right) \widehat{\delta \rho_1} + \nu \sum_{\sigma \ne 1} \widehat{\delta\rho_{\sigma}},
\end{equation}
and for the other states $\sigma\ne 1$
\begin{equation}
\partial_t \widehat{\delta \rho_\sigma} = \left( - {\bf k} \cdot {\frak D}_\sigma {\bf k} - i v {\bf k} \cdot {\bf e_\parallel} + \kappa \right) \widehat{\delta \rho_\sigma} - \mu \widehat{\delta \rho_1} - \frac{\kappa+\nu}{q-2} \sum_{\sigma' \ne 1,\sigma} \widehat{\delta \rho_{\sigma'}},
\end{equation}
where $\mu = M (q\beta J - 1 - 2\alpha M + \alpha M^2)$, $\nu = M [q\beta J - (q-1) + 2\alpha M + (q-1) \alpha M^2]$ and $\kappa = M [-(q-1)q \beta J + 1 +(q-4)\alpha M - \alpha M^2]$. The vector $R=\begin{pmatrix}
\widehat{\delta\rho_1} & \dots & \widehat{\delta\rho_q}
\end{pmatrix}^\intercal$ follows the equation $\partial_t R = M_{\rm liq} R$ with the matrix $M_{\rm liq}$ defined by its components
\begin{gather}
\left[M_{\rm liq}\right]_{1 \sigma} = \left(-{\bf k} \cdot {\frak D}_1 {\bf k} - i v {\bf k} \cdot {\bf e_\parallel} + (q-1) \mu \right) \delta_{1 \sigma} + \nu \left(1-\delta_{1 \sigma}\right), \nonumber \\
\label{defMliq}
\left[M_{\rm liq}\right]_{\sigma\ne1, \sigma'} = \left(-{\bf k} \cdot {\frak D}_\sigma {\bf k} - i v {\bf k} \cdot {\bf e_\parallel} + \kappa \right) \delta_{\sigma \sigma'} - \mu \delta_{1 \sigma'} - \frac{\kappa+\nu}{q-2} \left(1-\delta_{1 \sigma'} - \delta_{\sigma \sigma'}\right).
\end{gather}
To get the stability of the ordered homogeneous solutions, we need to calculate the eigenvalues of $M_{\rm liq}$. In the appendices \ref{appendixC2} and \ref{appendixD2}, we show for $q=4$ and $q=6$, respectively, that the ordered solution is stable only when the magnetization is equal to $M_0 + M_1 \delta$ corresponding to the largest value of $M$ in Eq.~(\ref{eqMag2}). For each of the two values of $q$, we can define two eigenvalues $\lambda_\parallel$ and $\lambda_\perp$: if $\lambda_\parallel<0$ a perturbation in the $x$-direction (the assumed direction of motion of the ordered homogeneous solution) is stable, and if $\lambda_\perp<0$ a perturbation in the $y$-direction (perpendicular to the assumed direction of motion) is stable. In appendix \ref{appendixC2}, for $q=4$, the expression of these two eigenvalues - via Eqs.~(\ref{lambdaPara4app}-\ref{lambdaPerp4app}) - are
\begin{gather}
\lambda_\parallel = -D + \frac{\mu+\nu}{3\mu-\nu} \frac{D\epsilon}{3} - \frac{4\mu[-3\mu^2+2\mu\nu+\nu(4\kappa+\nu)]}{(3\mu-\nu)^3(3\kappa+\nu)} \left(\frac{4D\epsilon}{3} \right)^2, \\
\lambda_\perp = -D - \frac{\mu+\nu}{3\mu-\nu} \frac{D\epsilon}{3} + \frac{4\mu}{(3\mu-\nu)(3\kappa+\nu)} \left(\frac{4D\epsilon}{3} \right)^2.
\end{gather}

\begin{figure}[t]
\begin{center}
\includegraphics[width=16cm]{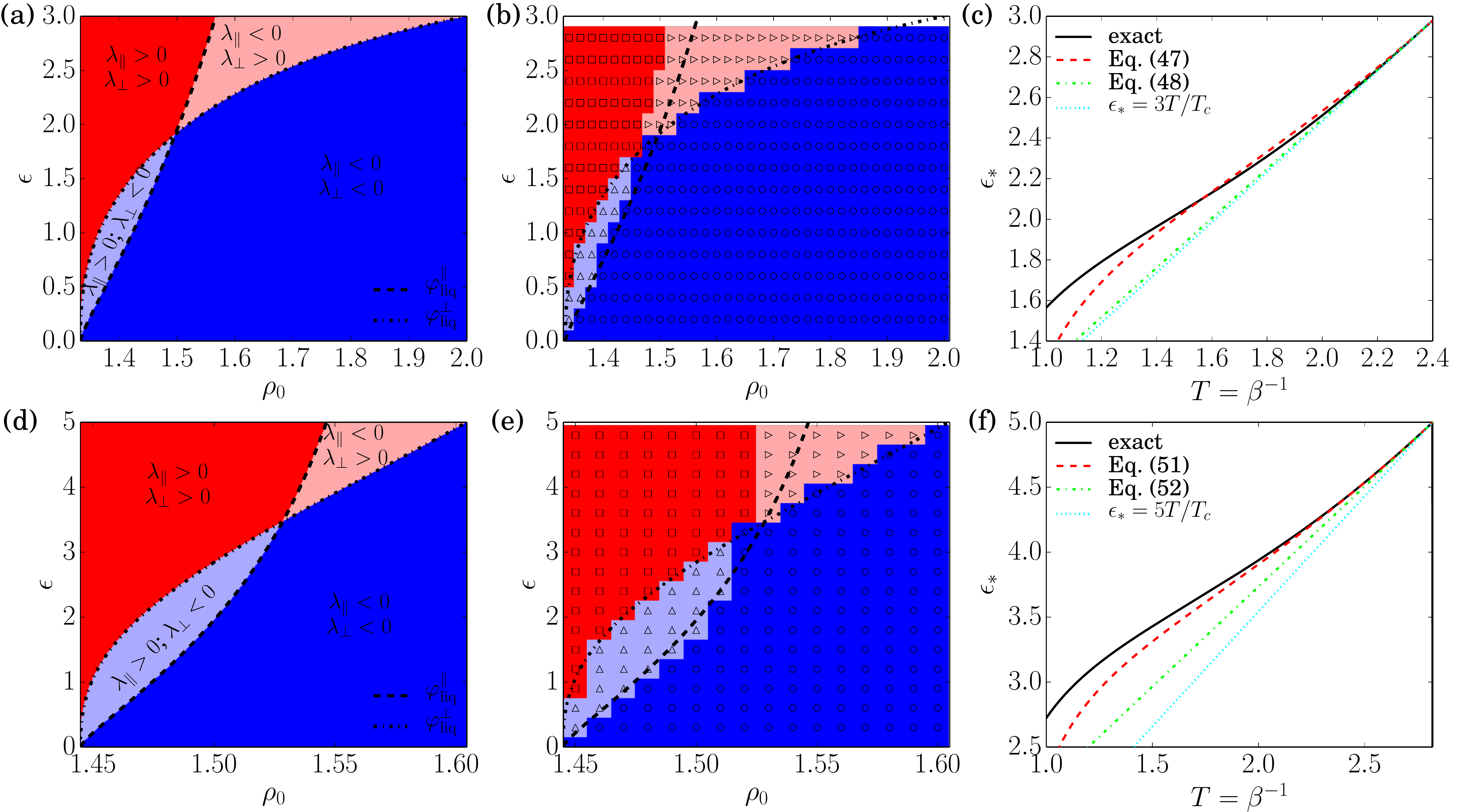}
\end{center}
\caption{(color online) (a-b) Stability diagrams for $q=4$ and $\beta=0.75$ and (d-e) for $q=6$ and $\beta=0.65$. In (a) and (d), the domains where a perturbation along $x$ ($\lambda_\parallel<0$) or along $y$ ($\lambda_\perp<0$) is stable, are displayed. The two stability domains are separated respectively by the spinodals $\varphi_{\rm liq}^\parallel$ and $\varphi_{\rm liq}^\perp$. In (b) and (e), these analytical domains are compared to numerical solutions, starting with a small perturbation around the ordered homogeneous solution. Disks: both perturbations vanish, up-triangles: only perturbation along $x$ grows, right-triangles: only perturbation along $y$ grows, squares: both perturbations grow. (c) $\epsilon_*$ value as a function of the temperature $T=\beta^{-1}$ for $q=4$. The same quantity is shown in (f) for $q=6$. Transverse and longitudinal stripes are stable respectively below and above this line. The plotted asymptotic expressions correspond to Eqs.~(\ref{epsilonstar4approx}-\ref{epsilonstar4leading}) in (c) and Eqs.~(\ref{epsilonstar6approx}-\ref{epsilonstar6leading}) in (f). \label{fig_APM_eigenvalues}}
\end{figure}

In Fig.~\ref{fig_APM_eigenvalues}(a) we show the velocity-density stability diagram for a fixed value of the temperature ($\beta=0.75$) plotted according to the signs of $\lambda_\parallel$ and $\lambda_\perp$. We can then extract the liquid spinodals $\varphi_{\rm liq}^\parallel(\epsilon)$ and $\varphi_{\rm liq}^\perp(\epsilon)$ as the boundaries of these domains, defined by the lines $\lambda_\parallel=0$ and $\lambda_\perp=0$, respectively. Note that $ \varphi_{\rm liq} = \max(\varphi_{\rm liq}^\parallel,\varphi_{\rm liq}^\perp)$ corresponds to the generalisation of the liquid spinodal defined in the AIM~\cite{ST2015}. In the Fig.~\ref{fig_APM_eigenvalues}(b), we represent the numerical behavior of an ordered homogeneous solution slowly perturbed along $x$ or $y$ directions, such that $\Delta \rho (t=0) = 10^{-3}$, obtained by solving numerically the Eqs.~(\ref{PDEhydro}) with FreeFem++. Four different regions are then derived depending on the evolution of the two kinds of perturbations. The analytical prediction of the spinodals agrees well with these regions, up to numerical inaccuracies in the region close to spinodals where the dynamics is slowed down.

The ordered homogeneous solution is always stable for a sufficiently large density (here $\rho_0>2$), whatever the direction of the perturbation. If the density $\rho_0$ is decreased at fixed value of $\epsilon$ and $\beta$, the liquid phase becomes unstable first for a perturbation along $x$ when $\epsilon<\epsilon_*$ and for a perturbation along $y$ when $\epsilon>\epsilon_*$ (here $\epsilon_*=2.0)$. The perturbation along $x$ and the perturbation along $x$ create a density profile $\rho(x,t)$ invariant in the $y$-direction and a density profile $\rho(y,t)$ invariant in the $x$-direction, respectively. Therefore we conclude that a transverse band is forming for small values of $\epsilon$ and a longitudinal lane for large values of $\epsilon$. Note that the stability of the phase-separated stripes is not changed whatever the value of the density $\rho_0$ between the two binodals $\rho_{\rm gas}$ and $\rho_{\rm liq}$, which plays a role for the volume fraction of liquid and gas. When the density $\rho_0$ is decreased (staying above the value $\rho_*$), the liquid phase becomes unstable for the two perturbations. The value of $\epsilon_*$, where the reorientation transition takes place, can be deduced from the equality of the spinodals $\varphi_{\rm liq}^\parallel = \varphi_{\rm liq}^\perp$ equivalent to the system $\lambda_\parallel=\lambda_\perp=0$. In the appendix \ref{appendixC2} we get an approximative expression of $\epsilon_*$ for a temperature close to $T_c$:
\begin{equation}
\label{epsilonstar4approx}
\epsilon_* = 3 \left[1+ \frac{16+23M_0}{40 M_0(-2+M_0+M_0^2)} M_1^2 + {\cal O}(M_1^4) \right],
\end{equation}
which gives at the first order in the $(T_c-T)$ expansion:
\begin{equation}
\label{epsilonstar4leading}
\epsilon_* \simeq 3 \left[1 - 0.981 \frac{T_c-T}{T_c} + \cdots \right].
\end{equation}
In Fig.~\ref{fig_APM_eigenvalues}(c), we compare these expression with the exact solution of $\epsilon_*$ solving numerically the system $\lambda_\parallel=\lambda_\perp=0$. Eq.~(\ref{epsilonstar4approx}) gives the best approximation and Eq.~(\ref{epsilonstar4leading}) is close to the line $\epsilon_*=3T/T_c$ also presented in Fig.~\ref{fig_APM_eigenvalues}(c). Below the line at $\epsilon=\epsilon_*$, the transverse bands are stable whereas the longitudinal lanes are obtained above this line. Note that at $\epsilon=3$ only longitudinal lanes are stable and the transverse bands are always observed at $\epsilon=0$.

In appendix \ref{appendixD2}, for $q=6$, the expression of two eigenvalues $\lambda_\parallel$ and $\lambda_\perp$ - via Eqs.~(\ref{lambdaPara6app}-\ref{lambdaPerp6app}) - are
\begin{gather}
\lambda_\parallel = -D + \frac{\mu+\nu}{5\mu-\nu} \frac{3D\epsilon}{10} - \frac{36\mu(-5\mu^2+2\mu\nu+\kappa\nu)}{(5\mu-\nu)^3(5\kappa+\nu)} \left(\frac{6D\epsilon}{5} \right)^2, \\
\lambda_\perp = -D - \frac{\mu+\nu}{5\mu-\nu} \frac{3D\epsilon}{10} + \frac{12\mu}{(5\mu-\nu)(5\kappa+\nu)} \left(\frac{6D\epsilon}{5} \right)^2.
\end{gather}
In the Fig.~\ref{fig_APM_eigenvalues}(d), we represent the velocity-density stability diagram for a fixed value of the temperature ($\beta=0.65$) plotted according to the sign of $\lambda_\parallel$ and $\lambda_\perp$. The different domains obtained for $q=4$ are retrieved and the spinodals are constructed at the same manner. In the Fig.~\ref{fig_APM_eigenvalues}(e), we reproduce, identically to the $q=4$ problem, the numerical behavior of an ordered homogeneous solution slowly perturbed along $x$ or $y$ directions solving numerically the Eqs.~(\ref{PDEhydro}) with FreeFem++, showing the accuracy of the spinodals. In the appendix \ref{appendixD2} we also obtain an approximative expression of $\epsilon_*$, where the reorientation transition takes place, for a temperature close to $T_c$:
\begin{equation}
\label{epsilonstar6approx}
\epsilon_* = 5 \left[1- \frac{4+9M_0}{12 M_0(1-M_0^2)} M_1^2 + {\cal O}(M_1^4) \right].
\end{equation}
which gives at the first order in the $(T_c-T)$ expansion:
\begin{equation}
\label{epsilonstar6leading}
\epsilon_* = \simeq 5 \left[1 - 0.869 \frac{T_c-T}{T_c} + \cdots \right].
\end{equation}
In Fig.~\ref{fig_APM_eigenvalues}(f), we represent the exact solution of $\epsilon_*$ solving numerically the system $\lambda_\parallel=\lambda_\perp=0$ and these two approximative results. As for $q=4$, Eq.~(\ref{epsilonstar6approx}) gives the best approximation, and transverse (resp. longitudinal) stripes are stable below (resp. above) this line.

\begin{figure}[t]
\begin{center}
\includegraphics[width=16cm]{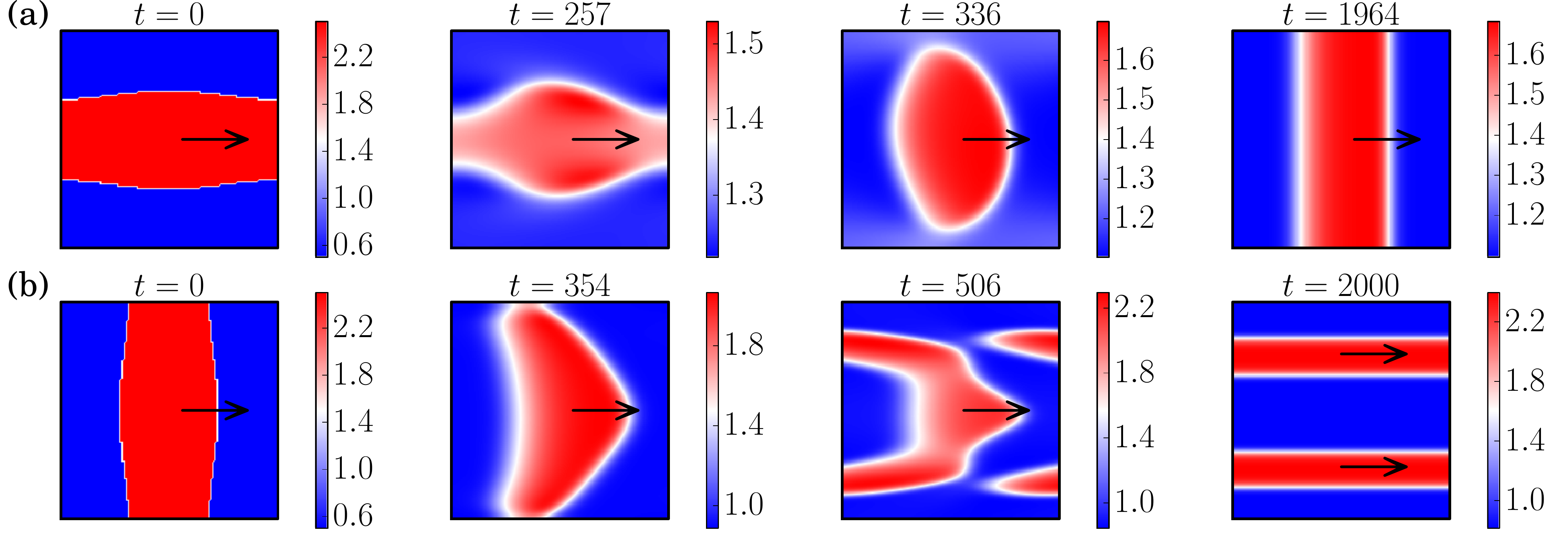}
\end{center}
\caption{(color online) Dynamics of the $q=4$ state APM for $\beta=0.75$, $\rho_0=1.33$ and $L=50$, and the numerical parameters $\Delta t= 0.2$ and ${\cal N}=100$. (a) The reorientation transition starting from an unstable longitudinal lane at $\epsilon=0.5<\epsilon_*$ transforming into a stable transverse band in the stationary state. (b) Opposite situation starting from an unstable transverse band at $\epsilon=2.5>\epsilon_*$ transforming into a stable longitudinal lane in the stationary state.\label{fig_APM_dynamics}}
\end{figure}

In Fig.~\ref{fig_APM_dynamics} we show the dynamics of the $q=4$ state APM for $\beta=0.75$, $\rho_0=1.33$ and two values of $\epsilon$: $\epsilon=0.5$ and $\epsilon=2.5$, below and above the reorientation transition at $\epsilon=2.0$, for which the stationary state is a transverse band and a longitudinal lane, respectively. In Fig.~\ref{fig_APM_dynamics}(a), a longitudinal lane is taken as an initial state and we observe the reorientation to a stable transverse band. In Fig.~\ref{fig_APM_dynamics}(b), the opposite situation is observed with a transverse band as the initial state, for which two stable longitudinal lanes are created. The number of stripes in one periodic square generally depends on the initial condition, but does not impact the values of the binodals $\rho_{\rm gas}$ and $\rho_{\rm liq}$ and the orientation of the stripes.


\section{Monte Carlo simulations on discrete lattices}
\label{section5}

In this section, we further investigate the numerical simulation results of $q=4$ and $q=6$ state APM. The models are respectively simulated on a square lattice and a triangular lattice of linear size $L=200$ with periodic boundary conditions applied on all sides. Simulations are performed for various control parameters: $\gamma=1$ and $D=1$ are kept constant throughout the simulations, $\beta = 1/T$ regulates the noise in the system, $\rho_0$ = $N/L^2$ defines the average particle density, and self-propulsion parameter $\epsilon$ dictates the effective velocity the particles: $\epsilon=q-1$ signifies complete self-propulsion whereas $\epsilon=0$ means pure diffusion. Starting from either a homogeneous or a semi-ordered initial condition, the Monte Carlo algorithm (Sec. \ref{section2}) evolves the system under various control parameters until the stationary distribution is reached. Following this, measurements are carried out and thermally averaged data are recorded. Note that, due to symmetries in APM, the phase separation occurs along the self-propulsion directions. 

A standard procedure that Monte Carlo simulation adopts for systems undergoing phase ordering kinetics is a random initial configuration. Nevertheless, in the current simulation, we have taken the initial configurations as semi-ordered (stripes of high and low densities) and verified that the final results are independent of the initial choice within the parameter regime. The advantage of a semi-ordered initial configuration over random initial configuration is that the former accelerates reaching the equilibrium. 

\subsection{Simulation results for $q = 4$ state APM}

\begin{figure}[t]
\centering
\includegraphics[width=16cm]{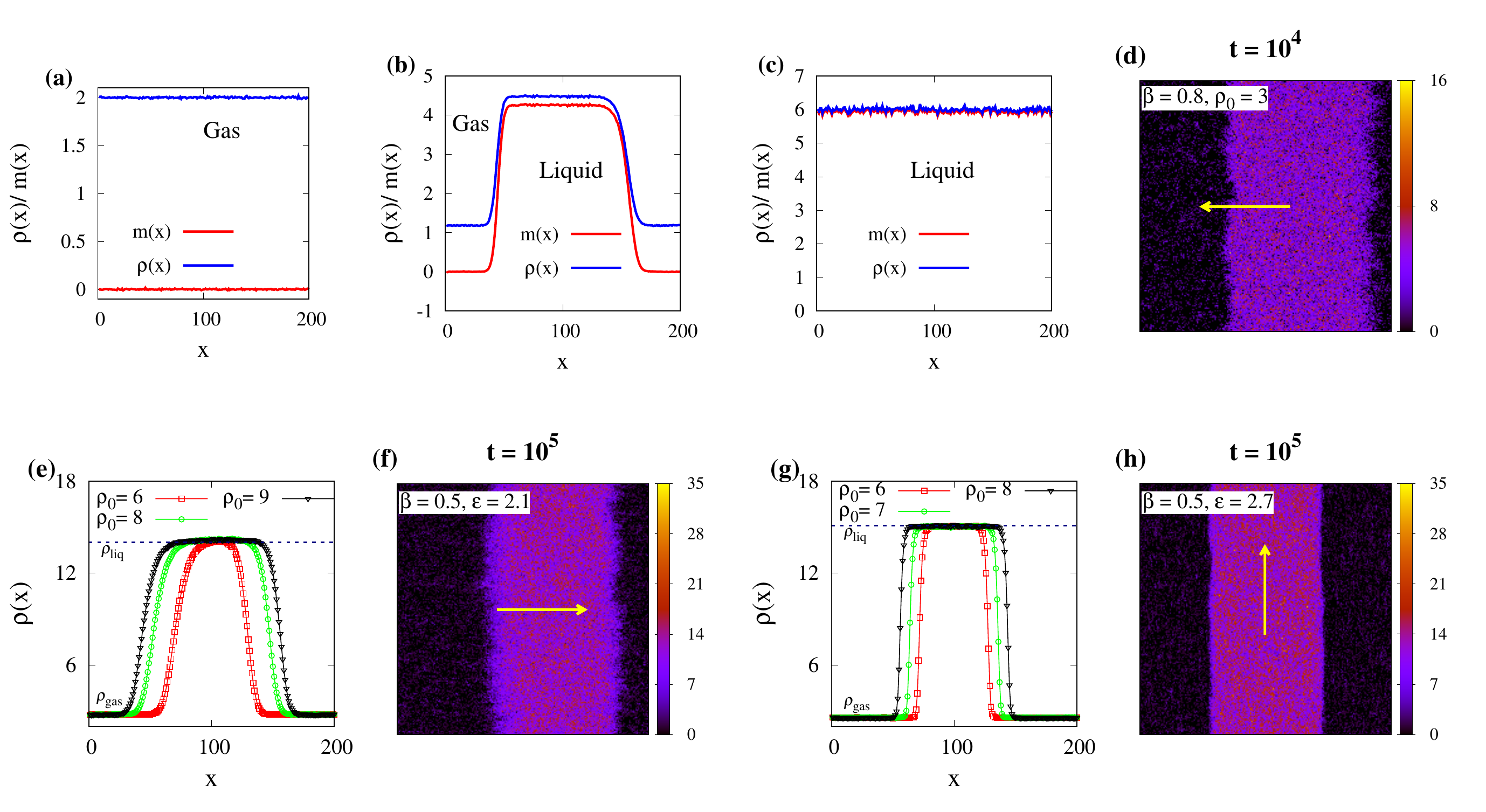}
\caption{(color online) (a--d) The three different phases of the 4-state APM for $\epsilon=0.9$. In (a), a disordered gaseous state evolves at $\beta=0.5$, $\rho_0=2$ while, in (c) an ordered liquid phase found at $\beta=0.95$, $\rho_0=6$. (b) At intermediate $\beta=0.8$, $\rho_0=3$, a stable liquid-gas coexistence phase is observed. (d) Stationary snapshot at $t=10^4$, corresponding to the data in (b), showing a transversely moving band with $\sigma=3$. (e--h) Phase separated density profiles of liquid ($\rho_{\rm liq}$) and gaseous ($\rho_{\rm gas}$) phases for 4-state APM at a fixed $\beta$ with increasing initial average $\rho_0$. (e) Density profiles for $\beta=0.5$ and $\epsilon=2.1$, with phase separated snapshot in (f) for $\rho_0=8$. (g) Density profiles for $\beta=0.5$ and $\epsilon=2.7$, and corresponding snapshot for $\rho_0=8$ is shown in (h). Arrows and color bars in (d,f,h) respectively represent direction of particles within the liquid stripes and on site particle density.}
\label{fig1}
\end{figure}

In order to identify the different phases of the stationary state, the density $\rho_0$, the temperature $\beta$, and the self-propulsion parameter $\epsilon$ are varied systematically, as control parameters, for a fixed diffusion constant $D = 1$. Fig.~\ref{fig1}(a--d) shows three different phases of APM for $\epsilon=0.9$: (a) disordered gaseous phase at a relatively high temperature $\beta=0.5$ and low density $\rho_0=2$ where the system is homogeneous with average magnetization $\langle m_i^{\sigma=1} \rangle \sim 0$, (b) liquid-gas coexistence phase at intermediate density $\rho_0$ = 3 and temperature $\beta=0.8$, where a stripe of polar liquid propagate transversely on a disordered gaseous background and (c) an ordered liquid phase at low temperature $\beta=0.95$ and high density $\rho_0=6$ where average magnetization $\langle m_i^{\sigma=1} \rangle \neq 0$ and can be significantly large depending on the average particle density $\rho_0$. The stationary snapshot in Fig.~\ref{fig1}(d), corresponding to the data in Fig.~\ref{fig1}(b), shows a polar liquid stripe moving transversely (denoted by arrow) and is predominantly constituted by particles with internal state $\sigma=3$.

In Fig.~\ref{fig1}(e--h), segregated density profiles of the liquid-gas coexistence phase are shown for $\beta=0.5$. Fig.~\ref{fig1}(e) and Fig.~\ref{fig1}(g), obtained for $\epsilon=2.1$ and $\epsilon=2.7$ respectively, suggest the broadening of the width of the polar liquid stripe with $\rho_0$ while keeping the densities of the liquid ($\rho_{\rm liq}$) and the gaseous ($\rho_{\rm gas}$) phases constant. The corresponding magnetization profiles (not shown here) show similar stripe with $0 < m_{\rm liq} < \rho_{\rm liq}$ and $m_{\rm gas} = 0$. The snapshots presented in Fig.~\ref{fig1}(f) and Fig.~\ref{fig1}(h) respectively correspond to Fig.~\ref{fig1}(e) and Fig.~\ref{fig1}(g) but for $\rho_0=8$. The interesting feature emerges from these snapshots is the directional switching of stripe propagation at higher $\epsilon$. We observe that the transverse orientation of the stripe with respect to the predominant drift of particles (represented by arrow) at $\epsilon=2.1$ becomes longitudinal at $\epsilon=2.7$. This is a novel feature of the APM.

\begin{figure}[t]
\centering
\includegraphics[width=16cm]{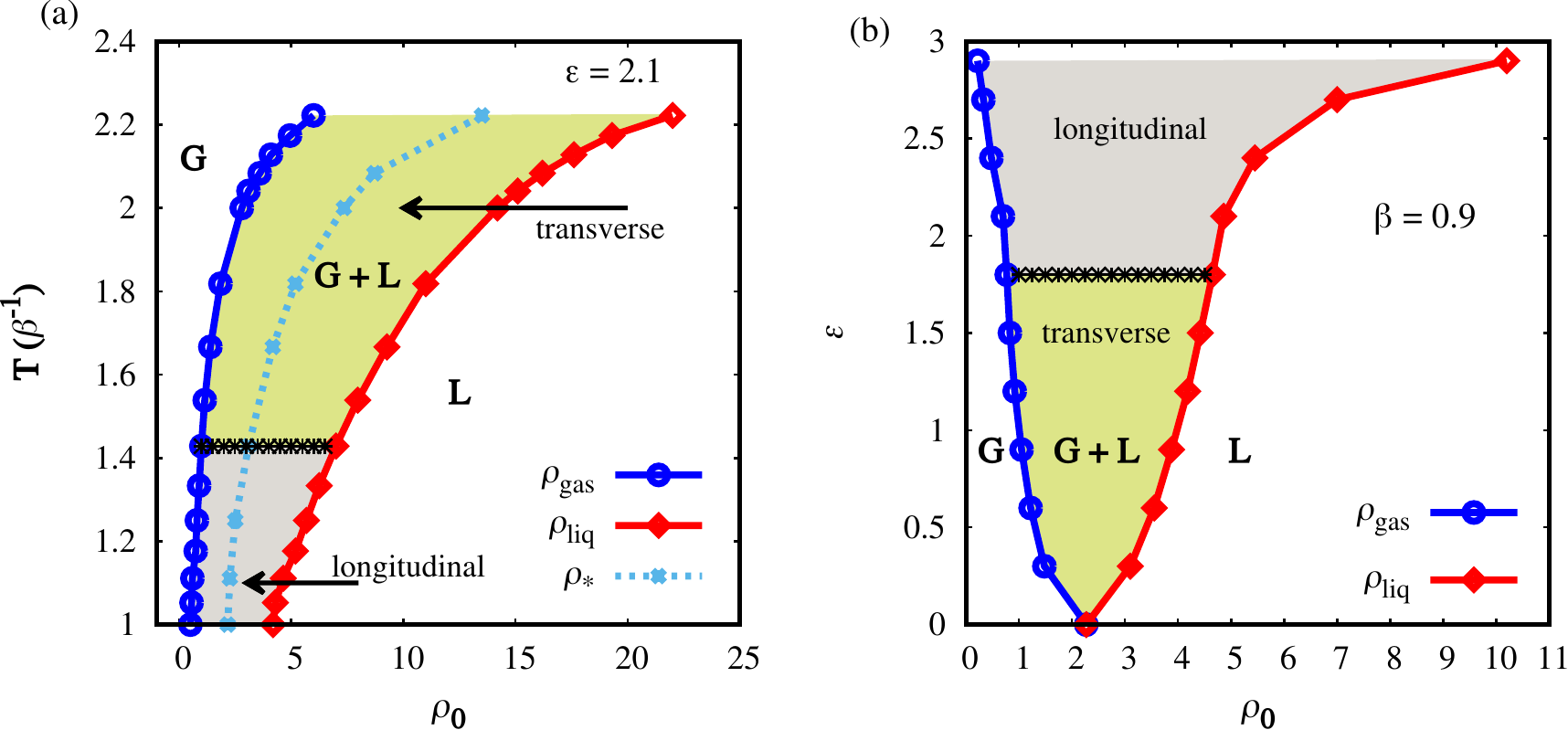}
\caption{(color online) (a--b) Phase diagrams of the $q=4$ state APM. (a) Temperature ($T$) versus density ($\rho_0$) phase diagram for fixed $\epsilon=2.1$, where $\rho_{\rm gas}$ and $\rho_{\rm liq}$ separate the three phases, gas (G), gas-liquid coexistence (G+L) and liquid (L). The dotted line in the G+L region indicates $\epsilon=0$ transition density ($\rho_*$) line as a function of the temperature. (b) $\epsilon$ versus $\rho_0$ phase diagram for $\beta=0.9$. The black dotted lines in (a) and (b) represent the reorientation transitions.} 
\label{fig3}
\end{figure}

Fig.~\ref{fig3}(a) represents the phase diagram in the $(T,\rho_0)$ plane for fixed $\epsilon=2.1$ where the binodals $\rho_{\rm gas}$ and $\rho_{\rm liq}$ lines delimit the gaseous ($G$), gas-liquid co-existence ($G+L$), and liquid ($L$) phases. For a fixed $\beta$, $\rho_{\rm gas}$ and $\rho_{\rm liq}$ are computed from the time averaged phase separated density profiles. The dashed line represents the density $\rho_*$ where the ordered-disordered transition occurs at $\epsilon=0$ which manifests a direct gas-liquid phase transition without going through a co-existence regime. $\epsilon$ versus $\rho_0$ phase diagram for a fixed temperature $\beta=0.9$ is shown in Fig.~\ref{fig3}(b). Once again, $\rho_{\rm gas}$ and $\rho_{\rm liq}$ lines separate the three phases and merges at a single point for $\epsilon=0$, which is the transition density point $\rho_*(\beta=0.9,\epsilon=0) \simeq 2.27$. One of the most distinguished feature of the APM, the reorientation transition of the co-existence phase liquid stripe, is depicted in both Fig.~\ref{fig3}(a) and Fig.~\ref{fig3}(b) through two different color shades. In the $(T,\rho_0)$ phase diagram we find that the transverse band at higher $T$ (lower $\beta$) switches to a longitudinal lane at lower $T$ (higher $\beta$) and this reorientation transition happens at $T \simeq 1.43$ ($\beta \simeq 0.7$) (represented by black dotted line) for $\epsilon=2.1$. In the $(\epsilon,\rho_0)$ phase diagram this reorientation approximately happens at $\epsilon \simeq 1.8$ (represented by black dotted line) for $\beta=0.9$ where $\epsilon<1.8$ is characterized by transverse band motion whereas $\epsilon \geqslant 1.8$ is characterized by longitudinal lane formation.

\begin{figure}[t]
\centering
\includegraphics[width=16cm]{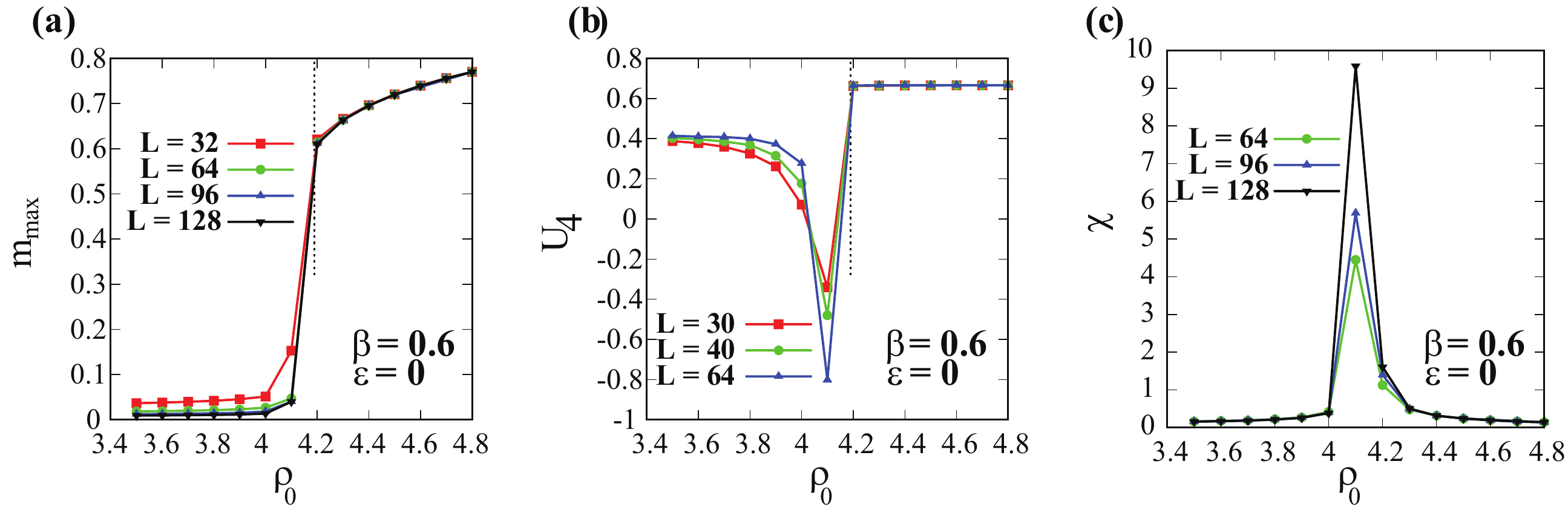}
\caption{(color online) Characterization of phase transition in the 4-state APM for $\epsilon=0$ and $\beta= 0.6$. (a) Maximal magnetization $m_{\rm max}$ versus $\rho_0$ for lattice size $L = 3$2, $64$, $96$, and $128$ are shown. The sudden jump in the magnetization signals a possible first-order phase transition. The phase transition is further corroborated by $U_4$ in (b), where the critical density $\rho_* = 4.19 \pm 0.01$ is estimated from the intersection of the data for various $L$. In (a) and (b), the point of transition is marked by the dotted lines. (c) Susceptibility ($\chi$) versus $\rho_0$ shows a discontinuous peak around $\rho_*$. } 
\label{fig4}
\end{figure}

In a purely diffusive APM, where particles hop without any bias ($\epsilon=0$), phase transition occur sharply from a low density homogeneous phase to a high density ordered phase with no intermediate gas-liquid co-existence. Data presented in Fig.~\ref{fig4}(a) shows the magnetization profile against density $\rho_0$ for $\beta=0.6$ and $\epsilon=0$, where a jump in the magnetization occurs around the transition $\rho_0$. Among the four different magnetizations corresponding to four internal states, we consider the maximal one ($m_{\rm max}$) plotted against $\rho_0$. The discontinuity in Fig.~\ref{fig4}(a) becomes sharper with increasing system sizes. This discontinuous change of a large $m_{\rm max}$ at a high density ($\rho_0>\rho_*$) ordered phase to a small $m_{\rm max} = 0$ at a relatively lower density ($\rho_0<\rho_*$) indicates the possibility of a first-order transition. Ideally, a fully ordered state acquires magnetization $m_{\rm max} \simeq 1$, however the ordered liquid phase suggests that all the particles on a lattice site may not belong to the same internal state and one realizes this from Eq.~(\ref{mag}) that $m_{\rm max} <1$. In Fig.~\ref{fig4}(b), we show the fourth-order cumulant of the magnetization (Binder cumulant) $U_4 = 1-\langle m_{\rm max}^4 \rangle/3 \langle m_{\rm max}^2 \rangle^2$ \cite{binder} versus $\rho_0$ across the transition and from the intersection of the $U_4$ curves for different $L$, we quantified the critical density $\rho_* (\beta=0.6,\epsilon=0) = 4.19 \pm 0.01$. Just below the transition density $\rho_*$, $U_4(L)$ becomes negative and falls further (approaches $-\infty$) with the increasing system size. A similar feature is again reflected in the susceptibility $\chi = \beta (\langle m_{\rm max}^2 \rangle-\langle m_{\rm max} \rangle^2)$ versus $\rho_0$ plot in Fig.~\ref{fig4}(c) where peaks are observed around the $\rho_*$ regardless of the system sizes. 

An important observation made here is that $\epsilon=0$ critical point of APM does not fall in the same universality class as the standard $q=4$ state Potts model with nearest-neighbor interactions. Solon \etal \cite{ST2015}, however, in their study of the AIM recover the $\epsilon=0$ critical point in the Ising universality class. For the $q$-state Potts model, it has been reported that the temperature-driven transitions are continuous for small $q \leqslant q_c$ and first-order for large $q > q_c$, with $q_c=4$ for the square-lattice with nearest-neighbor interactions and $q_c \simeq 2.8$ for the simple-cubic lattice \cite{baxter,wu,hartmann}. The reported critical exponents for the 4-state passive Potts model are $\beta^\prime=1/12$, $\gamma^\prime = 7/6$, and $\nu^\prime=2/3$ \cite{wu}, where $\beta^\prime$, $\gamma^\prime$, and $\nu^\prime$ are the critical exponents for magnetization, susceptibility, and correlation length, respectively. Finite-size scaling analysis, carried out with the data presented in Fig.~\ref{fig4} using these critical exponents does not yield any good data collapse and shows that the passive Potts model considered here with on-site interactions is different from the standard Potts model with nearest-neighbor interactions.

\subsection{Simulation results for $q = 6$ state APM}

\begin{figure}[t]
\centering
\includegraphics[width=16cm]{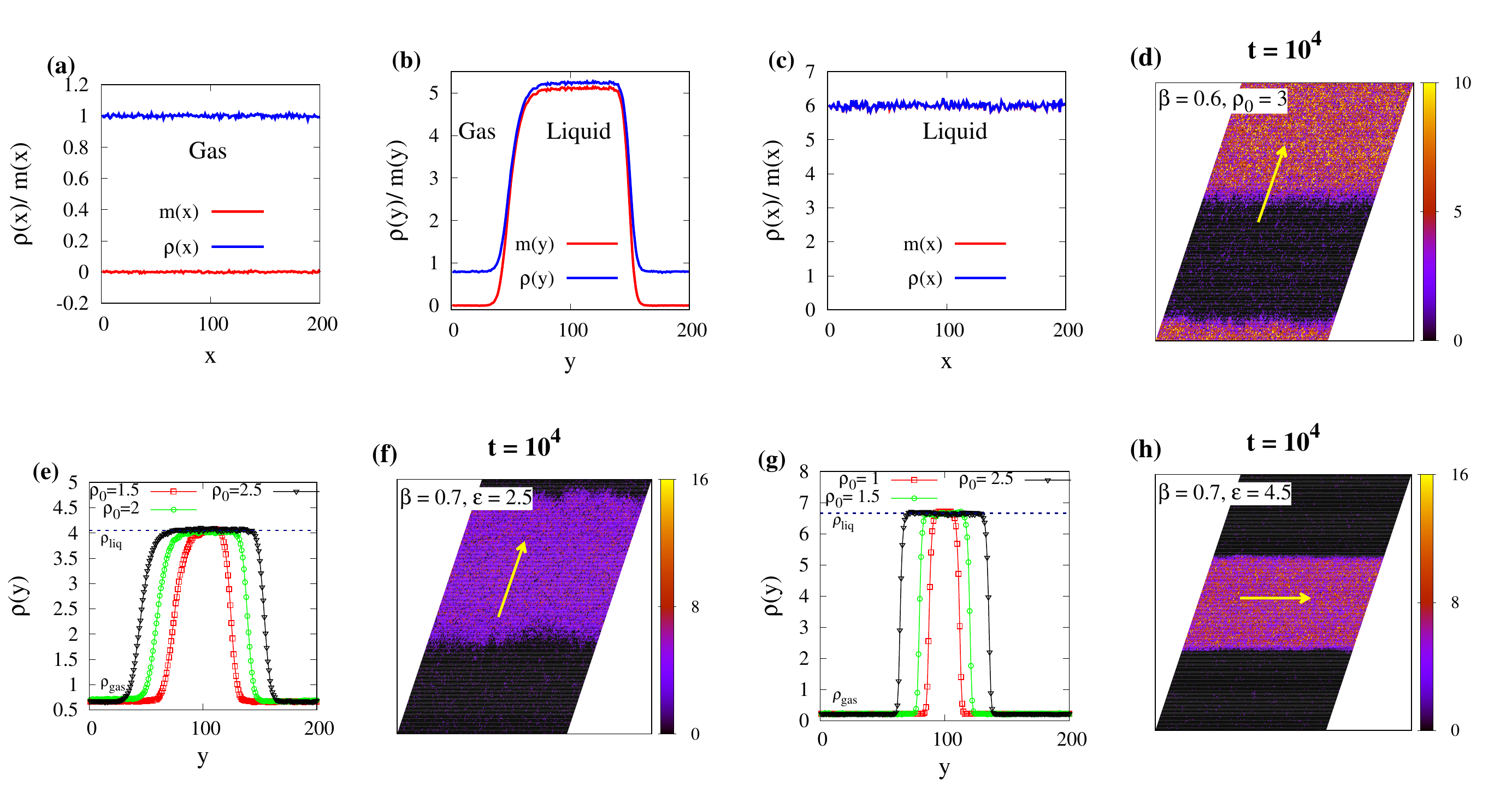}
\caption{(color online) (a--d) The three phases of 6-state APM for $\epsilon=2.5$. (a) Gas phase ($\beta=0.5$, $\rho_0=1$), (b) gas-liquid co-existence phase ($\beta=0.6$, $\rho_0=3$), and (c) liquid phase ($\beta=0.9$, $\rho_0=6$). The corresponding snapshot to (b) is shown in (d) where on site particle density is depicted in the color bar shown. (e--h) Phase separated density profiles for 6-state APM with $\beta=0.7$ and increasing initial average $\rho_0$. (e) Density profiles for $\epsilon=2.5$, and phase separated snapshot in (f) for $\rho_0=2.5$. (g) Density profiles for $\epsilon=4.5$, and corresponding snapshot for $\rho_0=2.5$ is shown in (h). Color bar represents on site particle density.} 
\label{fig5}
\end{figure}

The three different stationary phases of 6-state APM are shown in Fig.~\ref{fig5}(a--d) for $\epsilon=2.5$. A homogeneous gaseous phase for $\beta=0.5$ and $\rho_0=1$ with average magnetization $\langle m_i^{\sigma=5} \rangle \sim 0$ is shown in Fig.~\ref{fig5}(a), liquid-gas co-existence phase for $\beta=0.6$ and $\rho_0=3$ is shown in Fig.~\ref{fig5}(b), and Fig.~\ref{fig5}(c) shows the ordered liquid phase for $\beta=0.9$ and $\rho_0=6$. Please note that in Fig.~\ref{fig5}(c), although $m(x)<\rho(x)$, the magnetization profile is not distinctly visible as majority of particles on each site belong to $\sigma=1$ and therefore the difference of magnitude between $\rho(x)$ and $m(x)$ is very small. The snapshot corresponding to Fig.~\ref{fig5}(b) is shown in Fig.~\ref{fig5}(d) on a triangular lattice of linear size $L=200$. The transverse band displayed is primarily constructed by particles with $\sigma=2$.

Fig.~\ref{fig5}(e--h) demonstrates the phase-separated density profiles for 6-state APM. In Fig.~\ref{fig5}(e), the density profiles are shown for $\beta=0.7$, $\epsilon=2.5$, while varying the initial average density $\rho_0$. The nature of the segregated density profiles in this case are analogous to the density profiles for 4-state APM shown in Fig.~\ref{fig1}(e) where the width of the liquid fraction increases with increasing $\rho_0$ keeping the binodals $\rho_{\rm liq}$ and $\rho_{\rm gas}$ constant. Fig.~\ref{fig5}(f) corresponds to Fig.~\ref{fig5}(e) but for $\rho_0=2.5$ and shows transverse band motion along the predominant direction of the particles with $\sigma=2$. In Fig.~\ref{fig5}(g), the density profiles are shown for $\beta=0.7$, $\epsilon=4.5$, with varying initial average density $\rho_0$. Fig.~\ref{fig5}(h) displays the corresponding snapshot for $\rho_0=2.5$ and shows longitudinal lane formation along the predominant direction of the particles with $\sigma=1$.

\begin{figure}[t]
\centering
\includegraphics[width=16cm]{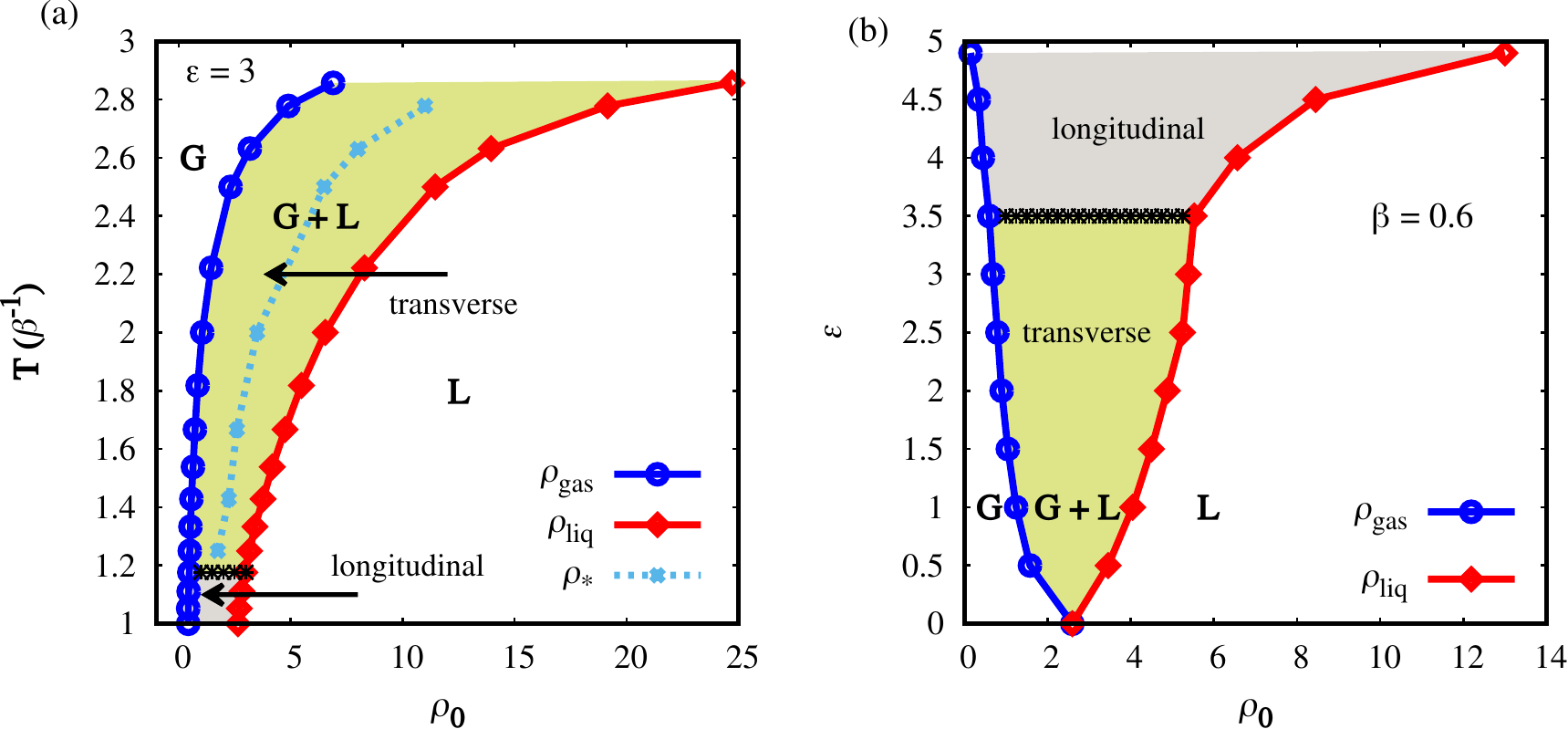}
\caption{(color online) Phase diagrams of the $q=6$-state APM, (a) $T$ versus $\rho_0$ for $\epsilon=3$ and (b) $\epsilon$ versus $\rho_0$ for $\beta=0.6$. $\rho_{\rm gas}$ and $\rho_{\rm liq}$ are the coexistence lines which respectively identify the densities of gas and liquid of the phase separated profiles. Analogous to Fig.~\ref{fig3}(a), $\rho_*$ in the G+L region of (a) indicates $\epsilon=0$ transition densities. The reorientation transitions from transverse to longitudinal motion of the liquid stripes as a function of $T$ and $\epsilon$ are indicated by black dotted lines in (a) and (b) respectively.} 
\label{fig7}
\end{figure}

The $(T,\rho_0)$ and $(\epsilon,\rho_0)$ phase-diagrams of the 6-state APM are presented in Fig.~\ref{fig7}(a) and Fig.~\ref{fig7}(b) respectively. Fig.~\ref{fig7}(a) is obtained for $\epsilon=3$ and analogous to Fig.~\ref{fig3}(a), the two co-existence lines $\rho_{\rm liq}$ and $\rho_{\rm gas}$ delimit the three different phases, $G$, $G+L$, and $L$. Notice that the critical point occurs at the temperature $T_c=3$ for 6-state APM as shown in Fig.~\ref{fig7}(a). The dashed line in the $G+L$ region indicates transition densities at $\epsilon=0$ through which a homogeneous gaseous phase can directly transform to a polar liquid phase with increasing $\rho_0$. The novel reorientation transition can also be observed in the 6-state APM. In Fig.~\ref{fig7}(a), a transition from low $\beta$ transverse motion to high $\beta$ longitudinal motion happens around $T \simeq 1.18$ ($\beta=0.85$) and is indicated by black dotted lines. A similar reorientation transition from low $\epsilon$ transverse motion to longitudinal motion at high $\epsilon$ happens at around $\epsilon \simeq 3.5$ in the $(\epsilon,\rho_0)$ phase diagram (Fig.~\ref{fig7}(b)) which has been constructed for $\beta=0.6$. Similar to Fig.~\ref{fig3}(b), here also the two binodals nicely separates the three phases, $G$, $G+L$, and $L$ and merges at the transition density point $\rho_*(\beta=0.6,\epsilon=0) \simeq 2.6$.

\begin{figure}[t]
\centering
\includegraphics[width=16cm]{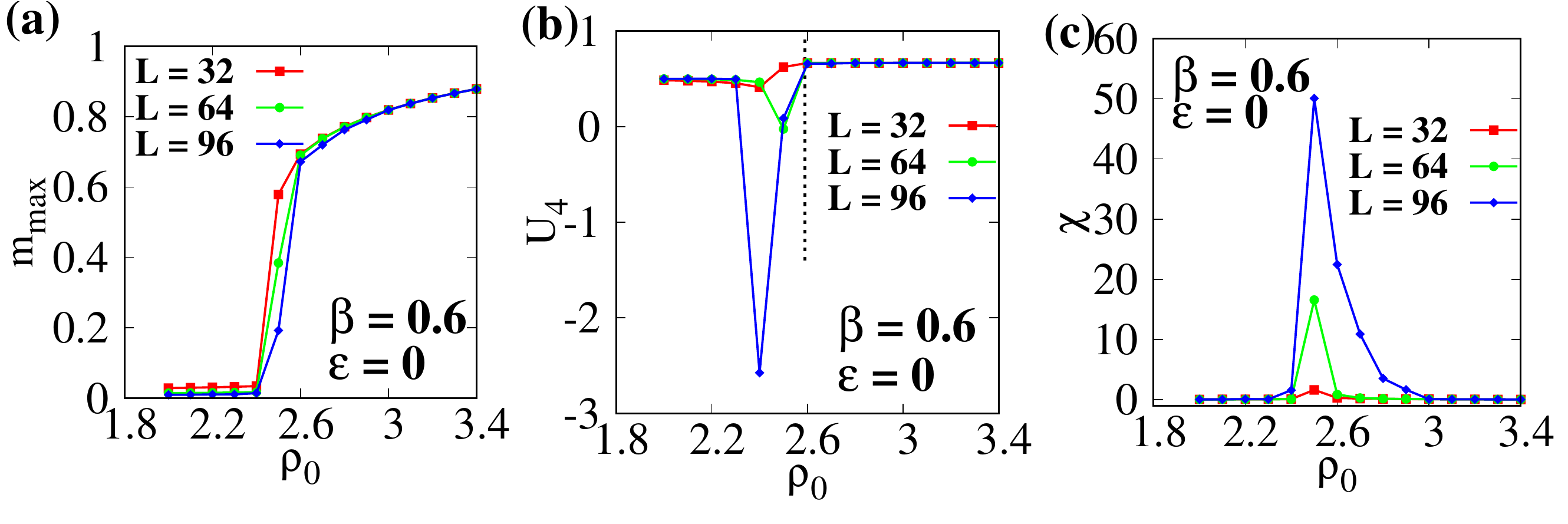}
\caption{(color online) Analogous to Fig.~\ref{fig4} but for $q=6$. (a) $m_{max}$ versus $\rho_0$ for $L$ = 32, 64, and 96. (b) Quantification of $\rho_*$ from $U_4$ versus $\rho_0$, where $\rho_*=2.60 \pm 0.02$. The dotted line in (b) mark the point of intersection of various $L$. The signature of a first-order phase transition in (a) and (b) are further validated in (c) where susceptibility ($\chi$) against $\rho_0$ shows a discontinuous peak around $\rho_*$.} 
\label{fig8}
\end{figure}

Fig.~\ref{fig8} shows the $\epsilon=0$ scenario for 6-state APM. The simulations are done on a triangular lattice with linear system sizes $L=32$, $64$, and $96$ and data presented are averaged over time and ensemble. $m_{\rm max}$ versus $\rho_0$ is presented in Fig.~\ref{fig8}(a) and is characterized by the discontinuous jump around the transition density and therefore identify the transition as a first-order phase transition \cite{baxter,wu,hartmann}. The plot of Binder cumulant $U_4$ against $\rho_0$ shown in Fig.~\ref{fig8}(b) quantify the transition density at $\rho_* (\beta=0.6,\epsilon=0) = 2.60 \pm 0.02$. The discontinuous peaks near around $\rho_*$ in the plot of $\chi$ versus $\rho_0$ further support the fact that the transition is first-order. 


\section{Summary and Discussion}
\label{section6}
In this study, we have characterized the flocking transition of the two-dimensional $q$-state APM and focused on the cases $q=4$ and $q=6$. Our study reveals that this phenomenon could be best described as a liquid-gas phase transition via a co-existence phase where a dynamic stripe of polar liquid evolves on a gaseous background, similar to what happens in the active Ising model (AIM) \cite{ST2013,ST2015,ST2015-2}. We explored the flocking transitions in the APM using three control parameters: the temperature $T = \beta^{-1}$, the average particle density $\rho_0$, and the hopping velocity $\epsilon$. We showed that, akin to the AIM \cite{ST2013,ST2015,ST2015-2}, the APM also has a phase transition from a high-temperature low-density gaseous phase with average magnetization $\langle m \rangle = 0$, to a low-temperature high-density polar liquid phase with average magnetization $0 < \langle m \rangle < \rho_0$ via a liquid-gas coexistence phase at intermediate densities and temperatures, where a part of the system consists of a liquid stripe moving through a disordered gaseous background. We further quantify the densities of the gas and liquid phases in the liquid-gas coexistence region and calculate the phase diagrams for $q=4$ and $q=6$. The homogeneous solution of the hydrodynamic equation can be derived analytically and give an expression for the average magnetization in terms of the average density $\rho_0$ and for the critical temperature $T_c$.

We found a novel reorientation transition of the phase-separated profiles from transversal band motion at low velocities and high temperatures to longitudinal lane formation at high velocities and low temperatures. The physical origin of this reorientation transition is the vanishing of the transverse diffusion constant for large velocities, stabilizing the longitudinal lane formation. A linear stability analysis of the homogeneous solutions, leading to an explicit equation for the spinodal lines, allows us to determine the velocity at the reorientation transition $\epsilon_*$ and to derive an analytical expression close to the critical temperature $T_c$. All predictions of the hydrodynamic theory were confirmed by Monte Carlos simulations of the microscopic model.

We further investigated the $\epsilon=0$ critical point, where the system undergoes a phase transition from a high density ordered phase to a low-density disordered phase at a critical value of the density $\rho_*$. The discontinuous jump of the average magnetization, obtained for both numerical simulations and coarse-grained hydrodynamic theory, indicates a first-order transition \cite{giordano}. This notion is further supported by the fourth-order Binder cumulant exhibiting a minimum around $\rho_*$. The minimum value tends to fall further (approaching $-\infty$) with increasing system sizes, a signature of the first-order transition \cite{kb,kb2}. The characteristics of the susceptibility ($\chi$) also suggest a first-order like phase transition. Nevertheless, unlike the AIM, $\epsilon=0$ critical point does not recover the standard $q$-state Potts model universality in our passive ($\epsilon=0$) Potts model. 

As a future perspective, it will be interesting to study the $q\to\infty$ limit of the APM. It should be noted that one does not expect to recover the Vicsek model in the $q\to\infty$ limit of the APM since the transition rules we introduced allows spin flips to an arbitrary new direction of motion, whereas the Vicsek model only allows small velocity changes in small time intervals. A better candidate of a lattice model with discrete velocity directions reproducing the Vicsek model in the $q\to\infty$ limit would be the $q$-state active clock model (ACM) in which larger direction changes are penalized by smaller transition probabilities (due to larger energy differences in the ferromagnetic alignment Hamiltonian). The ACM and its $q\to\infty$ limit will be studied in a forthcoming publication.

Another extension of our study would be to introduce a restriction on the maximum number of particles allowed on a single lattice site or to consider soft-core on-site interactions penalizing an increasing number of particles on a single site (akin to the well known soft-core Bose-Hubbard model). Preliminary studies of such "restricted" APMs indicate substantial differences in the stripe formation (Paul {\it et al.}, unpublished).

\section{Acknowledgments}
M.M. and H.R. were financially supported by the German Research Foundation (DFG) within the Collaborative Research Center SFB 1027. S.C. thanks the Indian Association for the Cultivation of Science, Kolkata for financial support. R.P. thanks CSIR, India, for support through Grant No. 03(1414)/17/EMR-II and the SFB 1027 for supporting his visit to the Saarland University for discussion and finalizing the project. 

\appendix


\section{Hydrodynamic limit of the Master equation}
\label{appendixA}

We define $a$ to be the lattice space between two sites which is small in the hydrodynamic limit ($L \gg 1$): $a \simeq 1/L \ll 1$. We define the continuous density of particles in the state $\sigma$ as $\rho_\sigma({\bf y},t) = n_i^\sigma(t)$ at the coordinate ${\bf y} \equiv (i_x, i_y) a$, where $(i_x, i_y)$ are the Cartesian coordinates of the former site $i$. The linear system size is then $l= (L-1) a$. Let ${\bf e_p}$ be the unitary vector in the direction $p$ with Cartesian coordinates ${\bf e_p}=(\cos \phi_p, \sin \phi_p)$ where $\phi_p=2 \pi (p-1)/q$ is the angle of the direction $p \in [1,q]$. The Taylor expansion of $\rho_\sigma({\bf y} + a {\bf e_p},t) $ gives
\begin{equation}
n_{i+p}^\sigma = \rho_\sigma({\bf y} + a {\bf e_p},t) = \rho_\sigma({\bf y},t) + a \frac{\partial \rho_\sigma}{\partial p} ({\bf y},t) + \frac{a^2}{2} \frac{\partial^2 \rho_\sigma}{\partial p^2} ({\bf y},t) + {\cal O}(a^3).
\end{equation}
for the derivative $ \frac{\partial}{\partial p} = {\bf e_p} \cdot \nabla_{\bf y}$. We can then expand the two terms in the Master equation (\ref{MasterEq}) involving the particle number difference on neighboring sites:
\begin{equation}
n_{i-\sigma}^\sigma - n_i^\sigma = - a \frac{\partial \rho_\sigma}{\partial \sigma}({\bf y},t) + \frac{a^2}{2} \frac{\partial^2 \rho_\sigma}{\partial \sigma^2} ({\bf y},t) + {\cal O}(a^3),\label{expConv}
\end{equation}
and
\begin{equation}
\sum_{p=1}^q \left[ n_{i-p}^\sigma - n_i^\sigma \right] = \sum_{p=1}^q \left[ a \frac{\partial \rho_\sigma}{\partial p} ({\bf y},t) + \frac{a^2}{2} \frac{\partial^2 \rho_\sigma}{\partial p^2} ({\bf y},t) \right] =
\begin{dcases}
a^2 \frac{q}{4} \nabla_{\bf y}^2 \rho_\sigma({\bf y},t) + {\cal O}(a^3) \qquad &q>2\\
a^2 \frac{q}{2} \nabla_{\bf y}^2 \rho_\sigma({\bf y},t) + {\cal O}(a^3) \qquad &q \le 2
\end{dcases},
\end{equation}
where we have used the following relations for an arbitrary function $F(x,y)$ in two dimensions
\begin{equation}
\sum_{p=1}^q \frac{\partial F}{\partial p}(x,y) = \sum_{p=1}^q {\bf e_p} \cdot {\bf e_x} \frac{\partial F}{\partial x}(x,y) + \sum_{p=1}^q {\bf e_p} \cdot {\bf e_y} \frac{\partial F}{\partial y}(x,y) = 0
\end{equation}
and
\begin{align}
\sum_{p=1}^q \frac{\partial^2 F}{\partial p^2}(x,y) &= \sum_{p=1}^q ({\bf e_p} \cdot {\bf e_x})^2 \frac{\partial^2 F}{\partial x^2}(x,y) + 2 \sum_{p=1}^q ({\bf e_p} \cdot {\bf e_x}) ({\bf e_p} \cdot {\bf e_y}) \frac{\partial^2 F}{\partial x \partial y}(x,y) + \sum_{p=1}^q ({\bf e_p} \cdot {\bf e_y})^2 \frac{\partial^2 F}{\partial y^2}(x,y) \nonumber\\
&= \begin{dcases} \frac{q}{2} \frac{\partial^2 F}{\partial x^2}(x,y) + \frac{q}{2} \frac{\partial^2 F}{\partial y^2}(x,y) = \frac{q}{2} \nabla^2 F(x,y) \qquad &q>2\\
q \frac{\partial^2 F}{\partial x^2}(x,y) \qquad &q \le 2.
\end{dcases}
\end{align}

For $q=2$, the master equation (\ref{MasterEq}) is equivalent to the Eq.~(15) of Ref.~\cite{ST2015} and it can be simplified as
\begin{equation}
\label{MasterEqAIM}
\partial_t \langle n_i^\sigma \rangle = D \left( \langle n_{i-1}^\sigma \rangle + \langle n_{i+1}^\sigma \rangle - 2 \langle n_i^\sigma \rangle \right) + \sigma D \epsilon \left( \langle n_{i-1}^\sigma \rangle - \langle n_{i+1}^\sigma \rangle \right) + \left\langle n_i^{-\sigma} W_{\rm flip}(-\sigma,\sigma) - n_i^\sigma W_{\rm flip}(\sigma,-\sigma) \right\rangle,
\end{equation}
where $\sigma=\pm 1$ are the one dimensional values of spins. Eq.~\eqref{MasterEqAIM} in the hydrodynamic limit takes the following form
\begin{equation}
\partial_t \langle \rho_\sigma \rangle = D \partial_x^2 \langle \rho_\sigma \rangle - \sigma v \partial_x \langle \rho_\sigma \rangle + \left\langle \rho_{-\sigma} W_{\rm flip}(-\sigma,\sigma) - \rho_\sigma W_{\rm flip}(\sigma,-\sigma) \right\rangle,
\end{equation}
Note that the diffusivity does not dependent on $\epsilon$, and this equation is equivalent to the Eqs.~(18-19) of Ref.~\cite{ST2015}.

For $q>2$, the Master equation (\ref{MasterEq}) becomes
\begin{equation}
\label{eqHydro000}
\partial_t \langle \rho_\sigma \rangle = \frac{q D}{4} \left(1 - \frac{\epsilon}{q-1} \right) a^2 \nabla_{\bf y}^2 \langle \rho_\sigma \rangle + \frac{q D \epsilon}{2(q-1)} a^2 \partial_\sigma^2 \langle \rho_\sigma \rangle - \frac{q D \epsilon}{q-1} a \partial_\sigma \langle \rho_\sigma \rangle + \sum_{\sigma' \ne \sigma } I_{\sigma \sigma'},
\end{equation}
where $I_{\sigma \sigma'}=\left\langle \rho_{\sigma'} W_{\rm flip}(\sigma',\sigma) - \rho_\sigma W_{\rm flip}(\sigma,\sigma') \right\rangle$ is the flipping term. Defining the new variable ${\bf x} = {\bf y}/a$, the linear system size is $l/a = L-1 \simeq L$ in the large system size limit $L \gg 1$ and the hydrodynamic equation (\ref{eqHydro000}) rewrites as
\begin{equation}
\label{eqHydro00}
\partial_t \langle \rho_\sigma \rangle = \frac{q D}{4} \left(1 - \frac{\epsilon}{q-1} \right) \nabla_{\bf x}^2 \langle \rho_\sigma \rangle + \frac{q D \epsilon}{2(q-1)} \partial_\parallel^2 \langle \rho_\sigma \rangle - \frac{q D \epsilon}{q-1} \partial_\parallel \langle \rho_\sigma \rangle + \sum_{\sigma' \ne \sigma } I_{\sigma \sigma'}.
\end{equation}
where $\partial_\parallel = {\bf e_\parallel} \cdot \nabla_{\bf x}$ is the derivative in the parallel direction ${\bf e_\parallel} = (\cos \phi_\sigma, \sin \phi_\sigma)$, with $\phi_\sigma = 2\pi(\sigma-1)/q$ the angle in the direction $\sigma$. Using the rotational invariance of the Laplacian $\nabla^2 = \partial_\parallel^2 + \partial_\perp^2$ with $\partial_\perp = {\bf e_\perp} \cdot \nabla_{\bf x}$, the derivative in the perpendicular direction ${\bf e_\perp} = (\sin \phi_\sigma, -\cos \phi_\sigma)$, Eq.~(\ref{eqHydro00}) can be rewritten as
\begin{equation}
\label{expAppendixA}
\partial_t \langle \rho_\sigma \rangle = D_\parallel \partial_\parallel^2 \langle \rho_\sigma \rangle + D_\perp \partial_\perp^2 \langle \rho_\sigma \rangle - v \partial_\parallel \langle \rho_\sigma \rangle + \sum_{\sigma' \ne \sigma } I_{\sigma \sigma'}
\end{equation}
where $D_{\parallel} = qD[1+\epsilon/(q-1)]/4$ and $D_{\perp} = qD[1-\epsilon/(q-1)]/4$ are the diffusion constants in the parallel direction ${\bf e_\parallel}$ and perpendicular direction ${\bf e_\perp}$, respectively, and $v=qD\epsilon/(q-1)$ is the self-propulsion velocity in the parallel direction ${\bf e_\parallel}$.


\section{Expansion of the flipping term: refined mean-field equations}
\label{appendixB}

We consider the flipping term defined before as
\begin{equation}
I_{\sigma \sigma'} = \left\langle \rho_{\sigma'} W_{\rm flip}(\sigma',\sigma) - \rho_\sigma W_{\rm flip}(\sigma,\sigma') \right\rangle,
\end{equation}
where $\rho_\sigma$ is the continuous version of $n_i^\sigma$. From the definition of $W_{\rm flip}(\sigma,\sigma')$, given by Eq.~(\ref{defWflip}), this flipping term writes
\begin{equation}
\label{fliptermApp0}
I_{\sigma \sigma'} = \gamma \left\langle \rho_{\sigma'} \exp\left[-\frac{q\beta J}{\rho}\left(\rho_{\sigma'} - \rho_\sigma-1\right)\right] - \rho_\sigma \exp\left[-\frac{q\beta J}{\rho}\left(\rho_\sigma - \rho_{\sigma'}-1\right)\right] \right\rangle.
\end{equation}
We can then factorize the flipping term by $\gamma \langle \exp(q\beta J/\rho) \rangle \simeq \gamma \exp(q\beta J/\rho_0)$ using the mean-field approximation, which will be set to $1$ in the following simplification. Note that this multiplicative constant does not change the value of the homogeneous solutions. Eq.~(\ref{fliptermApp0}) then becomes:
\begin{equation}
\label{refIij}
I_{\sigma \sigma'} \simeq \left\langle \rho_{\sigma'} \exp\left[-\frac{q\beta J}{\rho}(\rho_{\sigma'}-\rho_{\sigma})\right] - \rho_{\sigma} \exp\left[-\frac{q\beta J}{\rho}(\rho_{\sigma}-\rho_{\sigma'})\right] \right\rangle.
\end{equation}

\begin{figure}[t]
\begin{center}
\includegraphics[width=8cm]{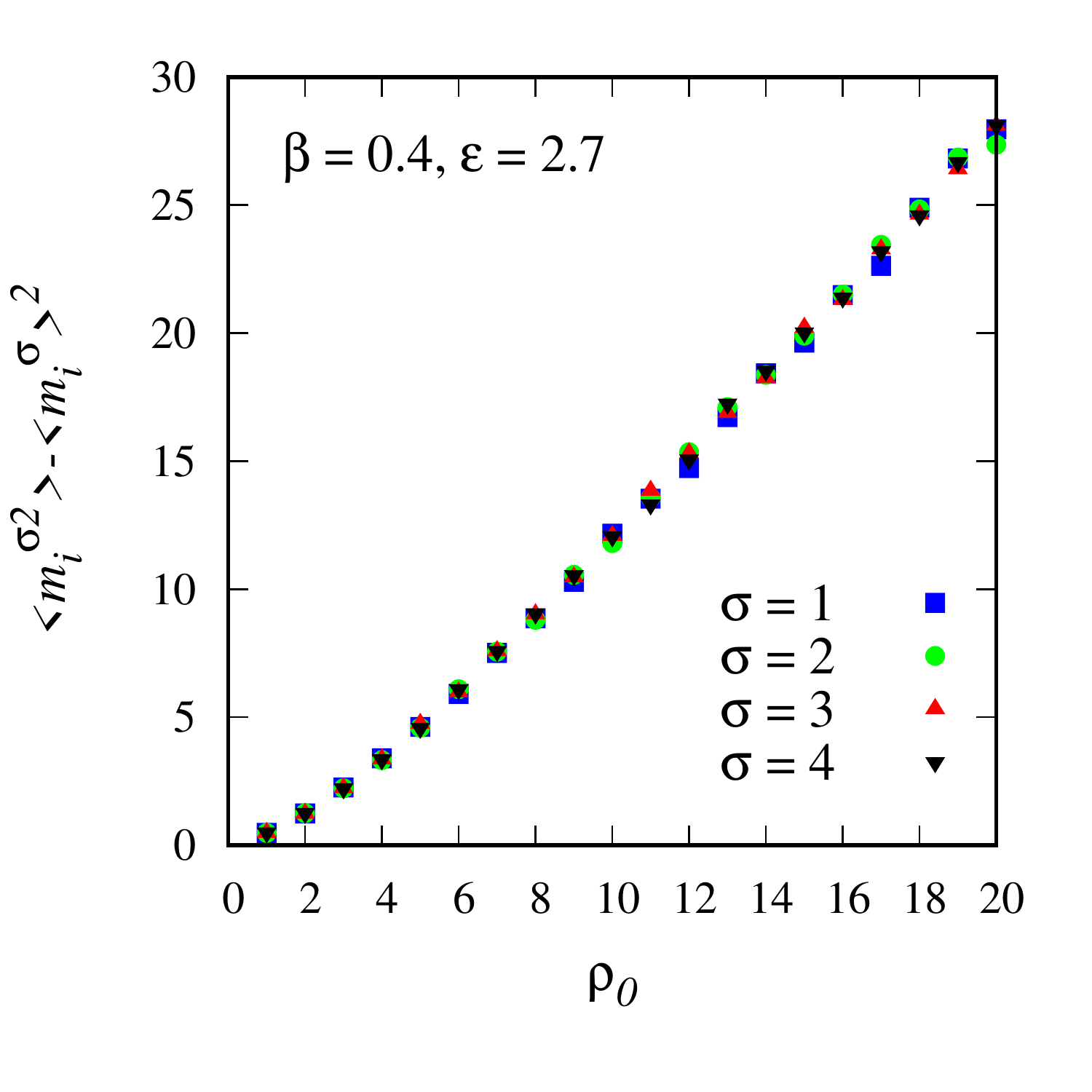}
\caption{Evolution of the variance $\langle {m_i^\sigma}^2 \rangle - \langle {m_i^\sigma \rangle}^2$ of the magnetization $m_\sigma$ with the mean population $\langle \rho_i \rangle = \rho_0$ in the disordered phase ($q=4$ and $\beta=0.4$). The relation is linear and identical for all states $\sigma$.} \label{figAPMalpha}
\end{center}
\end{figure}

The mean-field (MF) expression of Eq.~(\ref{refIij}) never shows stable phase-separated profiles and always predicts the trivial homogeneous solution. As for the AIM \cite{ST2015}, the simple mean-field approximation fails to predict the results for the microscopic model. Following \cite{ST2015} we derive a refined MF hydrodynamic equation for the particle density $\rho_\sigma$ including the first order of fluctuations in the magnetization, defined by Eq.~(\ref{mag}) such that $m_\sigma = (q\rho_\sigma-\rho)/(q-1)$ is assumed small compared to the particle density $\rho$. Then the quantity
\begin{equation}
\label{rel1}
\rho_{\sigma'}-\rho_{\sigma} = \frac{q-1}{q}(m_{\sigma'}-m_{\sigma})
\end{equation}
is small compared to the particle density $\rho$ and Eq.~(\ref{refIij}) can be expanded as
\begin{equation}
\label{fliptermExpand}
I_{\sigma \sigma'} \simeq \left\langle (\rho_{\sigma'}-\rho_{\sigma}) - \frac{q\beta J}{\rho}(\rho_{\sigma'}-\rho_{\sigma})(\rho_{\sigma'}+\rho_{\sigma}) + \frac{(q\beta J)^2}{2\rho^2} (\rho_{\sigma'}-\rho_{\sigma})^3 - \frac{(q\beta J)^3}{6\rho^3} (\rho_{\sigma'}-\rho_{\sigma})^3 (\rho_{\sigma'}+\rho_{\sigma}) + \dots \right\rangle.
\end{equation}
Using (\ref{rel1}) and the following relation
\begin{equation}
\label{rel2}
\rho_{\sigma'}+\rho_{\sigma} = \frac{2\rho}{q} + \frac{q-1}{q}(m_{\sigma'}+m_{\sigma}),
\end{equation}
we get the r.h.s of (\ref{fliptermExpand}) up to the order $(m_\sigma-m_{\sigma'})^3$:
\begin{equation}
\label{Iij2}
I_{\sigma \sigma'} \simeq \left\langle (2\beta J-1) \xi (m_{\sigma}-m_{\sigma'}) + \frac{q\beta J}{\rho} \xi^2 (m_{\sigma}-m_{\sigma'}) (m_{\sigma}+m_{\sigma'}) - \alpha \xi^3 \frac{(m_{\sigma}-m_{\sigma'})^3}{\rho^2} + \dots \right\rangle,
\end{equation}
where $\xi=(q-1)/q$ and $\alpha= (q \beta J)^2(1-2\beta J/3)/2$. We assume that all magnetization $m_\sigma$ are identically distributed Gaussian variables with the variance $\alpha_m \langle \rho \rangle$ proportional to the local mean population, with mean values linked by the relation $\langle m_r \rangle + \langle m_l \rangle + \langle m_u \rangle + \langle m_d \rangle=0$. In Fig.~\ref{figAPMalpha}, we verify this approximation by MC simulations of the microscopic model in the disordered state ($q=4$, $\beta=0.4$). Expanding the average value in Eq.~(\ref{Iij2}) and assuming that $m_\sigma$ and $\rho$ are uncorrelated, we obtain
\begin{equation}
I_{ij} \simeq (2 \beta J-1) \xi \langle m_{\sigma} - m_{\sigma'} \rangle + \frac{q \beta J}{\langle \rho \rangle} \xi^2 \langle m_{\sigma}^2 - m_{\sigma'}^2 \rangle-\alpha \xi^3 \frac{\langle (m_{\sigma}-m_{\sigma'})^3 \rangle}{\langle \rho \rangle^2}.
\end{equation}
Using the Gaussian distribution properties, we can show that the second moment is $\langle m_{\sigma}^2 \rangle = \langle m_{\sigma} \rangle^2 + \alpha_m \langle \rho \rangle$ and the third moment is $\langle (m_{\sigma}-m_{\sigma'})^3 \rangle = \langle m_{\sigma}- m_{\sigma'} \rangle^3+6 \alpha_m \langle m_{\sigma} - m_{\sigma'} \rangle\langle \rho \rangle$. The expression of $I_{\sigma \sigma'}$ then becomes
\begin{equation}
I_{\sigma \sigma'} \simeq \left[2\beta J-1 - \frac{6\alpha_m \alpha \xi^2}{\langle \rho \rangle}\right]\xi \langle m_{\sigma} - m_{\sigma'} \rangle + \frac{q \beta J}{\langle \rho \rangle} \xi^2 \langle m_{\sigma} - m_{\sigma'} \rangle \langle m_{\sigma} + m_{\sigma'} \rangle -\alpha \xi^3 \frac{ \langle m_{\sigma}- m_{\sigma'} \rangle^3}{\langle \rho \rangle^2}.
\end{equation}

Using the averaged value of Eqs.~(\ref{rel1}) and~(\ref{rel2}), the flipping term then rewrites
\begin{equation}
\label{expAppendixB}
I_{\sigma \sigma'} = \left[\frac{q\beta J}{\langle \rho \rangle} \langle \rho_{\sigma} + \rho_{\sigma'} \rangle -1 - \frac{r}{\langle \rho \rangle} - \alpha \frac{\langle \rho_{\sigma}-\rho_{\sigma'}\rangle^2}{\langle \rho \rangle ^2}\right]\langle \rho_{\sigma}-\rho_{\sigma'}\rangle
\end{equation}
where $r= 6(q-1)^2\alpha_m \alpha/q^2$ is a new positive parameter depending only on the temperature, since $\alpha_m$ is a unknown function of $\beta$.


\section{The limit $q \rightarrow + \infty$}
\label{appendixInfty}

In Sec.~\ref{section2} the flipping and hopping rates are given by Eqs.~(\ref{defWflip}) and (\ref{defWhop}). Since the total hopping rate $q D$ and the maximal value of $\epsilon$ must stay finite, we rescale these microscopic parameters such that $\overline D = q D$ and $\overline \epsilon = \epsilon/(q-1)$. In the $q\to\infty$ limit, each particle has a continuous spin-state $\phi \in [0,2\pi]$. The density of particles $n_i(\phi)$ in the state $\phi=2\pi(\sigma-1)/q$ is then related to the number of particle $n_i^\sigma$ in the state $\sigma$ by
\begin{equation}
n_i(\phi) = \lim_{q \rightarrow + \infty} \frac{q}{2\pi} n_i^\sigma.
\end{equation}
The number of particle $n_i^\sigma$ decays as $1/q$ since the number of particle on site $i$ scales as $\rho_0$ (the average density) which is independent of $q$. This definition gives the equivalence for the total density of particles
\begin{equation}
\rho_i = \int_0^{2\pi} d\phi \ n_i(\phi) = \sum_{\sigma=1}^q n_i^\sigma
\end{equation}
for the Riemann integral defined for all continuous function $f(x)$ in $[0,2\pi]$ such that
\begin{equation}
\int_0^{2\pi} dx \ f(x) = \lim_{q\rightarrow + \infty} \sum_{\sigma=1}^q \frac{2\pi}{q} f(x_\sigma).
\end{equation}

According to Eq.~(\ref{defWhop}) a particle in state $\phi$ jumps during the time interval $\Delta t$ in a random direction in $[\phi',\phi'+d\phi']$ with probability $qD[1-\epsilon/(q-1)] d\phi' \Delta t/2\pi$ or rate $\overline D(1- \overline \epsilon)$ and in the direction $\phi$ with probability $qD\epsilon \Delta t/(q-1)$ or rate $\overline D \overline \epsilon$. From Eq.~(\ref{defWflip}), a particle in state $\phi$ performs a flip during the time interval $\Delta t$ to a state in $[\phi',\phi'+d\phi']$ with probability
\begin{equation}
\frac{q\gamma}{2\pi} \exp \left(\frac{q \beta J}{\rho_i}\right) \exp\left\{-\frac{2 \pi \beta J}{\rho_i} \left[ n_i(\phi) - n_i(\phi') \right]\right\} d\phi' \Delta t,
\end{equation}
which is equivalent to the flipping rate density for a spin flip from state $\phi$ to state $\phi'$:
\begin{equation}
W_{\rm flip}(\phi,\phi') = \frac{\overline \gamma}{2\pi} \exp\left\{-\frac{2\pi \beta J}{\rho_i} \left[n_i(\phi) - n_i(\phi')\right]\right\}.
\end{equation}
where $\overline \gamma = q \gamma \exp(q \beta J/\rho_0)$ is chosen from the mean-field value of the prefactor.

In the hydrodynamic limit and the limit $q \rightarrow +\infty$, we define the density of particles in the state (direction) $\phi$ and at the 2d position ${\bf x}$ as
\begin{equation}
\rho({\bf x},t;\phi) = \lim_{q \rightarrow + \infty} \frac{q}{2\pi} \rho_\sigma({\bf x},t).
\end{equation}
After simplifications, the Eq.~(\ref{PDEhydro}) becomes
\begin{gather}
\frac{\partial \rho(\phi)}{\partial t} = \nabla_{\bf x} \cdot {\frak D}(\phi) \nabla_{\bf x} \rho(\phi) - v {\bf e_\parallel}(\phi) \cdot \nabla_{\bf x} \rho(\phi) \nonumber\\ 
+ \overline \gamma \int_0^{2\pi} \frac{d \phi'}{2\pi} \left[\frac{2\pi\beta J}{\rho}(\rho(\phi)+\rho(\phi')) -1 - \frac{r}{\rho} - \overline \alpha \frac{(\rho(\phi)-\rho(\phi'))^2}{\rho^2}\right](\rho(\phi)-\rho(\phi')),
\end{gather}
where $\overline \alpha= (2 \pi \beta J)^2(1-2\beta J/3)/2$, $v= \overline D \overline \epsilon$, ${\bf e_\parallel}(\phi) = (\cos \phi, \sin \phi)$ and
\begin{equation}
{\frak D}(\phi) = \frac{\overline D}{4} I_2 +
\frac{\overline D \overline \epsilon}{4}
\begin{pmatrix}
\cos 2\phi & \sin 2\phi \\
\sin 2\phi & - \cos 2\phi
\end{pmatrix}.
\end{equation}
Note that the magnetization is given by $m(\phi) = \rho(\phi) - \rho/2\pi$, which fulfills the identity $\int\limits_0^{2\pi} d\phi \ m(\phi) = 0$.

The disordered homogeneous solution is given by a constant density $\rho(\phi)= \rho_0/2\pi$. The ordered homogeneous solutions are deduced from the discrete ordered homogeneous solutions written as $\rho_\sigma = (\rho_0-m_0)/q + \delta_{\sigma 1} m_0$ for a polar ordered liquid in the $\sigma=1$ state. The density function of these ordered homogeneous solutions are then
\begin{equation}
\rho(\phi) = \frac{\rho}{2\pi} + \delta(\phi) m_0
\end{equation}
where $\delta(\phi)$ is zero everywhere except at $\phi=0$ with a value $1$. $m_0$ is the magnetization of state $\phi=0$ defined by
\begin{equation}
\frac{m_0}{\rho_0} = \frac{\pi \beta J}{\overline \alpha} \left\{ 1 \pm \sqrt{1+ \frac{\mu_0 \overline \alpha}{(\pi \beta J)^2}} \right\},
\end{equation}
with the previously defined quantity $\mu_0= 2 \beta J - 1 -r/\rho_0$. This expression is equivalent to Eq.~(\ref{eqMag}) in the $q\to\infty$ limit.


\section{Expression of the polarization vector}
\label{appendixPolarization}

The polarization vector ${\bf P}_i(t)$ is defined on site $i$ as the average direction of the self-propulsion:
\begin{equation}
{\bf P}_i(t) = \frac{1}{\rho_i} \sum_{k=1}^{\rho_i} {\bf e_i^k}(t)
\end{equation}
where ${\bf e_i^k}(t)$ is the direction of the $k$-th particle on site $i$ at the time $t$. This expression can be rewritten as ${\bf P}_i = {\bf v}_i / v$ where the average velocity on site $i$ is the ratio of the average displacement ${\bf \Delta X}_i$ occurring in $\Delta t$ by the corresponding time interval $\Delta t$. This average displacement can be calculated as
\begin{equation}
{\bf \Delta X}_i = \sum_{\sigma=1}^q {\bf e_\sigma} \mathbb{P}_i(\sigma)
\end{equation}
where ${\bf e_\sigma}$ is the unitary vector in the direction $\sigma$ with the angle $\phi_\sigma = 2\pi (\sigma-1)/q$ and $\mathbb{P}_i(\sigma)$ is the probability to move in the direction $\sigma$ on site $i$. From the hopping rate, Eq.~(\ref{defWhop}), this probability is equal to
\begin{equation}
\mathbb{P}_i(\sigma) = \frac{n_i^\sigma}{\rho_i} D(1+\epsilon) \Delta t + \frac{\rho_i - n_i^\sigma}{\rho} D\left(1-\frac{\epsilon}{q-1}\right) \Delta t
\end{equation}
by decomposing the motion for the two classes of particles: the first one when the direction $\sigma$ is favored (for particles with spin-state $\sigma$) and the second one when the direction $\sigma$ is not favored (for particles with spin-state different from $\sigma$). Using the expression of the magnetization given by Eq.~(\ref{mag}), we get the expression
\begin{equation}
\mathbb{P}_i(\sigma) = \left[ 1 + \epsilon \frac{m_i^\sigma}{\rho_i} \right] \Delta t.
\end{equation}
Since $\sum\limits_{p=1}^q {\bf e_\sigma} = 0$, the average velocity writes
\begin{equation}
{\bf v}_i = D \epsilon \sum_{\sigma=1}^q {\bf e_\sigma} \frac{m_i^\sigma}{\rho_i},
\end{equation}
and using the expression $v=qD\epsilon/(q-1)$ we obtain the expression of the polarization vector
\begin{equation}
\label{polarization0}
{\bf P}_i = \frac{q-1}{q}\sum_{\sigma=1}^\sigma {\bf e_\sigma} \frac{m_i^\sigma}{\rho_i} = \sum_{\sigma=1}^q {\bf e_\sigma} \frac{n_i^\sigma}{\rho_i}.
\end{equation}
This expression is compatible with the intuitive value of ${\bf P}_i$ in the gas phase (${\bf P}_i = {\bf 0}$) and when all particles are in the same state $\sigma$: ${\bf P}_i = {\bf e_\sigma}$.

In the hydrodynamic limit, the polarization vector writes ${\bf P}({\bf x},t) = \langle {\bf P}_i \rangle (t)$ at the 2d position ${\bf x}$. Using the mean-field approximation, Eq.~(\ref{polarization0}) becomes
\begin{equation}
{\bf P}({\bf x},t) = \frac{q-1}{q}\sum_{\sigma=1}^\sigma {\bf e_\sigma} \frac{m_\sigma({\bf x},t)}{\rho({\bf x},t)} = \sum_{\sigma=1}^q {\bf e_\sigma} \frac{\rho_\sigma({\bf x},t)}{\rho({\bf x},t)},
\end{equation}
which corresponds to Eq.~(\ref{polarization}) used in the main text.


\section{Linear stability analysis for $q=4$ state APM}

\subsection{Linear stability of disordered homogeneous solution}
\label{appendixC1}

For $q=4$, the symmetries of the problem between $x$ and $y$ direction implies that the perturbations along $x$ and $y$ axis on the disordered homogeneous solution are identical. So, without any loss of generality, we can consider here $k_y=0$. From the Eq.~(\ref{defMgas}), the matrix $M_{\rm gas}$ writes then
\begin{equation}
M_{\rm gas}^{(x)}=\left(\begin{matrix}
- D_\parallel k_x^2 - i k_x v + 3 \mu_0 & -\mu_0 & -\mu_0 & -\mu_0\\
-\mu_0 & - D_\perp k_x^2 + 3 \mu_0 & -\mu_0 & -\mu_0 \\
-\mu_0 & -\mu_0 & - D_\parallel k_x^2 + i k_x v + 3 \mu_0 & -\mu_0 \\
-\mu_0 & -\mu_0 & -\mu_0 & - D_\perp k_x^2 + 3 \mu_0
\end{matrix}\right).
\end{equation}
Up to order ${\cal O}(k_x^2)$, Mathematica \cite{Mathematica} gives the expressions of the four eigenvalues $\lambda_{\rm gas}^i$:
\begin{gather}
\lambda_{\rm gas}^1 = 4 \mu_0 -D_\perp k_x^2, \\
\lambda_{\rm gas}^2 = \left(-\frac{D_\parallel+D_\perp}{2} + \frac{v^2}{8\mu_0} \right) k_x^2 + \cdots , \\
\lambda_{\rm gas}^{3,4} = 4 \mu_0 \pm \frac{i\sqrt{2} v k_x}{2} - \left( \frac{3D_\parallel+D_\perp}{4} + \frac{v^2}{16 \mu_0} \right) k_x^2 + \cdots.
\end{gather}
Note that if $\lambda$ is an eigenvalue then its complex conjugate $\overline{\lambda}$ is also an eigenvalue. Since $D_\perp>0$ and $D_\parallel>0$, the real part of all eigenvalues are negative when $\mu_0<0$, defining the condition for a stable disordered homogeneous solution.

\subsection{Linear stability of ordered homogeneous solution}
\label{appendixC2}

For an ordered homogeneous solution, the ($x$, $y$) symmetry is broken, implying that a perturbation along $x$ axis has a different behavior that a perturbation along $y$ axis. The expression of the matrix $M_{\rm liq}$ is derived from Eq.~(\ref{defMliq}) with $\mu = M (4\beta J - 1 - 2\alpha M + \alpha M^2)$, $\nu = M (4\beta J - 3 + 2\alpha M + 3 \alpha M^2)$ and $\kappa = M (-12 \beta J + 1 - \alpha M^2)$ for $q=4$. Let us consider first a perturbation in the $x$ direction ($k_y=0$). The matrix then writes
\begin{equation}
M_{\rm liq}^{(x)}=\left(\begin{matrix}
- D_\parallel k_x^2 - i k_x v + 3 \mu & \nu & \nu & \nu\\
-\mu & - D_\perp k_x^2 + \kappa & -(\kappa+\nu)/2 & -(\kappa+\nu)/2 \\
-\mu & -(\kappa+\nu)/2 & - D_\parallel k_x^2 + i k_x v + \kappa & -(\kappa+\nu)/2 \\
-\mu & -(\kappa+\nu)/2 & -(\kappa+\nu)/2 & - D_\perp k_x^2 + \kappa
\end{matrix}\right),
\end{equation}
where $D_\parallel=D(1+\epsilon/3)$ and $D_\perp=D(1-\epsilon/3)$. Up to the order ${\cal O}(k_x^2)$, Mathematica gives the expression of the four eigenvalues $\lambda_{{\rm liq}, x}^i$:
\begin{gather}
\lambda_{{\rm liq}, x}^1 = \frac{3\kappa+\nu}{2} - D_\perp k_x^2, \\
\lambda_{{\rm liq}, x}^2 = \frac{3\kappa+\nu}{2} + \frac{2ivk_x}{3} - \left[ \frac{2D_\parallel+D_\perp}{3} + \frac{4(3\kappa-6\mu+\nu) v^2}{27(3\kappa+\nu)(\kappa-2\mu+\nu)} \right] k_x^2, \\
\lambda_{{\rm liq}, x}^3 = (3\mu-\nu) - \frac{i (9\mu+\nu) v k_x}{9\mu-3\nu} - \left[ \frac{9D_\parallel \mu -(D_\parallel+2D_\perp)\nu}{3(3\mu-\nu)} + \frac{4\nu(-36\kappa\mu +81 \mu^2 -42 \mu \nu +\nu^2) v^2}{27(3\mu-\nu)^3(\kappa-2\mu+\nu)} \right] k_x^2, \\
\lambda_{{\rm liq}, x}^4 = \frac{i(\mu+\nu)v k_x}{3\mu - \nu} - \left[ \frac{(D_\parallel + 2D_\perp) \mu - D_\parallel \nu}{3\mu - \nu} + \frac{4\mu(-3\mu^2+2\mu\nu+\nu(4\kappa+\nu)) v^2}{(3\mu-\nu)^3(3\kappa+\nu)} \right] k_x^2.
\end{gather}

Now we look at a perturbation in the $y$ direction ($k_x=0$). The matrix $M_{\rm liq}$ now writes
\begin{equation}
M_{\rm liq}^{(y)}=\left(\begin{matrix}
- D_\perp k_y^2 + 3 \mu & \nu & \nu & \nu\\
-\mu & - D_\parallel k_y^2 - i k_y v + \kappa & -(\kappa+\nu)/2 & -(\kappa+\nu)/2 \\
-\mu & -(\kappa+\nu)/2 & - D_\perp k_y^2 + \kappa & -(\kappa+\nu)/2 \\
-\mu & -(\kappa+\nu)/2 & -(\kappa+\nu)/2 & - D_\parallel k_y^2 + i k_y v + \kappa
\end{matrix}\right),
\end{equation}
where $D_\parallel=D(1+\epsilon/3)$ and $D_\perp=D(1-\epsilon/3)$. Up to the order ${\cal O}(k_y^2)$, the eigenvalues are
\begin{gather}
\lambda_{{\rm liq}, y}^{1,2} = \frac{3\kappa+\nu}{2} \pm \frac{ivk_y}{\sqrt{3}} - \left[ \frac{2D_\parallel + D_\perp}{3} + \frac{2(3\kappa-6\mu+\nu) v^2}{9(3\kappa+\nu)(\kappa-2\mu+\nu)} \right] k_y^2 + \cdots \\
\lambda_{{\rm liq}, y}^3 = (3\mu-\nu) - \left[ \frac{9D_\perp \mu - (2D_\parallel+D_\perp)\nu}{3(3\mu-\nu)} + \frac{4\nu v^2}{9(3\mu-\nu)^3(\kappa-2\mu+\nu)} \right] k_y^2 + \cdots \\
\lambda_{{\rm liq}, y}^4 = \left[ \frac{-(2D_\parallel + D_\perp) \mu + D_\perp \nu}{3\mu - \nu} + \frac{4\mu v^2}{(3\mu-\nu)(3\kappa+\nu)} \right] k_y^2 + \cdots.
\end{gather}

The ordered homogeneous solution is then stable if $3\kappa+\nu<0$ and $3\mu-\nu<0$ for the two different perturbations. Since $3\mu - \nu = 8M ( \beta J - \alpha M)$, the only stable ordered homogeneous solution satisfies $M>\beta J/ \alpha$. From Eq.~(\ref{eqMag2}), the magnetization of the stable solution is then equal to
\begin{equation}
\label{stableM}
M = \frac{\beta J}{\alpha} + \sqrt{\frac{r}{\alpha \rho_*}} \sqrt{\frac{\rho_0-\rho_*}{\rho_0}} = M_0 + M_1 \delta,
\end{equation}
where $\rho_*$ has been defined in Eq.~(\ref{defrhostar}); $M_0=\beta J / \alpha$ and $M_1 = \sqrt{r/\alpha \rho_*}$ are temperature dependent constants; and $\delta = \sqrt{(\rho_0-\rho_*)/\rho_0}$ is a variable with values between $0$ and $1$. Moreover, $3\kappa + \nu = 2M(-16 \beta J + \alpha M)$ implying that $M< 16 \beta J / \alpha$ to have a stable solution, which is always satisfied from the maximal value of M: $M<3\beta J/\alpha$ from Eq.~(\ref{stableM}).

However, the stability of the two different perturbations differs from $\lambda_{{\rm liq}, x}^4$ and $\lambda_{{\rm liq}, y}^4$. The perturbation along $x$ is stable only if
\begin{equation}
\label{lambdaPara4app}
\lambda_\parallel = \frac{\Re \lambda_{{\rm liq}, x}^4}{k_x^2} = -D + \frac{\mu+\nu}{3\mu-\nu} \frac{D\epsilon}{3} - \frac{4\mu[-3\mu^2+2\mu\nu+\nu(4\kappa+\nu)]}{(3\mu-\nu)^3(3\kappa+\nu)} \left(\frac{4D\epsilon}{3} \right)^2
\end{equation}
is negative and the perturbation along $y$ is stable only if
\begin{equation}
\label{lambdaPerp4app}
\lambda_\perp = \frac{\Re \lambda_{{\rm liq}, y}^4}{k_y^2} = -D - \frac{\mu+\nu}{3\mu-\nu} \frac{D\epsilon}{3} + \frac{4\mu}{(3\mu-\nu)(3\kappa+\nu)} \left(\frac{4D\epsilon}{3} \right)^2
\end{equation}
is negative. These two eigenvalues can be rewritten as
\begin{equation}
\label{defEIGEN}
\lambda_\parallel = -D + Q_1 \frac{D\epsilon}{3} + Q_3 \left(\frac{4D\epsilon}{3} \right)^2, \qquad
\lambda_\perp = -D - Q_1 \frac{D\epsilon}{3} + Q_2 \left(\frac{4D\epsilon}{3} \right)^2
\end{equation}
for the quantities $Q_i$ independent of $\epsilon$ defined by
\begin{equation}
Q_1 = \frac{\mu+\nu}{3\mu-\nu}, \qquad Q_2 = \frac{4\mu}{(3\mu-\nu)(3\kappa+\nu)}, \qquad Q_3 = - \frac{4\mu[-3\mu^2+2\mu\nu+\nu(4\kappa+\nu)]}{(3\mu-\nu)^3(3\kappa+\nu)}.
\end{equation}
From this stability analysis, we can remark that transverse bands will be formed when $\lambda_\perp < 0 < \lambda_\parallel$ whereas longitudinal lanes will be created in the opposite case $\lambda_\parallel < 0 < \lambda_\perp$. Then, the reorientation transition happens when $\lambda_\parallel = 0$ and $\lambda_\perp =0$, and from the Eq.~(\ref{defEIGEN}) the value of the drift $\epsilon_*$ at the reorientation transition is given by
\begin{equation}
\label{eqepsilonstar}
\frac{4D\epsilon_*}{3} = \sqrt{\frac{2D}{Q_2+Q_3}} = \frac{-Q_1}{2(Q_3-Q_2)},
\end{equation}
where the second equality defines the value of $\rho_0$ where the reorientation transition takes place, which can be rewritten as
\begin{equation}
\label{eqdeltastar}
-Q_1 = \frac{2\sqrt{2D}(Q_3-Q_2)}{\sqrt{Q_2+Q_3}}.
\end{equation}

With Eq.~(\ref{stableM}) we can now rewrite the quantities $Q_1$, $Q_2$ and $Q_3$ as a function of $\delta$ with coefficients depending only on $M_0$ and $M_1$. With Mathematica we obtain after simplifications
\begin{gather}
Q_1(\delta) = -M_0 - \frac{M_1}{2} \frac{1+\delta^2}{\delta},\\
Q_2(\delta) = \frac{M_0^2+2M_0-M_1^2}{(-15 M_0 + \delta M_1)(M_0 + \delta M_1)} \left[ - \frac{M_1}{4} \frac{1+\delta^2}{\delta} + \frac{1-M_0}{2}\right], \\
Q_3(\delta) = -Q_2(\delta) \left[ \frac{M_1}{8} \frac{1+\delta^2}{\delta} - \frac{15M_0}{8 \delta^2} - \frac{2+13 M_0}{8} - \frac{5M_0(3M_0+1)}{4\delta M_1} \right].
\end{gather}
Eq.~(\ref{eqdeltastar}) is then satisfied for $\delta=\delta_*$. Since the inversion of this equation to get the exact expression of $\delta_*$ is too complicated, we look at the solution close to the critical point: $T \rightarrow T_c$. For this limiting case, the ordered-disordered transition takes place for $\rho_* \gg 1$, implying that $M_1 \ll 1$. We can then look at a solution $\delta_*$ as an asymptotic expansion in $M_1$, such that $\delta_* = \delta_1 M_1 + \delta_2 M_1^2 + \delta_3 M_1^3 + \cdots$. Taking $D=1$, we get with Eq.~(\ref{eqdeltastar})
\begin{equation}
\delta_* = \frac{1}{2(1-M_0)} M_1 - \frac{1+2M_0}{16(M_0-1)^3(2+M_0)} M_1^3 + {\cal O}(M_1^5)
\end{equation}
and the quantities $Q_1$, $Q_2$ and $Q_3$ evaluated at $\delta= \delta_*$ are equal to
\begin{gather}
Q_1(\delta_*) = -1 + \frac{3 M_1^2}{8(-2+M_0+M_0^2)} + {\cal O}(M_1^4),\\
Q_2(\delta_*) = \frac{M_1^2}{80 M_0 (1-M_0)} + {\cal O}(M_1^4), \\
Q_3(\delta_*) = \frac{1}{8} - \frac{3(4+7M_0)M_1^2}{160 M_0(-2+M_0+M_0^2)} + {\cal O}(M_1^4).
\end{gather}
Thus, with Eq.~(\ref{eqepsilonstar}), we obtain the expression of $\epsilon_*$ where the reorientation transition happens:
\begin{equation}
\epsilon_* = 3 \left[1+ \frac{16+23M_0}{40 M_0(-2+M_0+M_0^2)} M_1^2 + {\cal O}(M_1^4) \right].
\end{equation}
In section \ref{section4}, we have shown that $T_c^{-1} = 1 - \sqrt{22}/8$. Then we get that
\begin{equation}
M_1^2 \simeq \frac{8}{63} (-143+32\sqrt{22}) \frac{T_c-T}{T_c}, \qquad M_0 = \frac{2}{21} (-5+2\sqrt{22}),
\end{equation}
leading to the expression of $\epsilon_*$ as
\begin{equation}
\epsilon_* = 3 \left[1 + \frac{3520-993\sqrt{22}}{1160} \frac{T_c-T}{T_c} + \cdots \right] \simeq 3 \left[1 - 0.981 \frac{T_c-T}{T_c} + \cdots \right] \simeq \frac{3T}{T_c} + 0.057 \frac{T_c-T}{T_c} + \cdots.
\end{equation}
So, we can approximate at the leading order that the reorientation transition happens at $\epsilon_* = 3 T/T_c$.


\section{Linear stability analysis for $q=6$ state APM}

\subsection{Linear stability of disordered homogeneous solution}
\label{appendixD1}

For $q=6$, the ($x$, $y$) symmetry is not present, implying that the perturbations along $x$ and $y$ axis on the disordered homogeneous solution are not identical. For a perturbation along $x$ ($k_y=0$), from Eq.~(\ref{defMgas}), the matrix $M_{\rm gas}$ writes
\begin{equation}
M_{\rm gas}^{(x)}=\left(\begin{smallmatrix}
- D_\parallel^x k_x^2 - i k_x v + 5 \mu_0 & -\mu_0 & -\mu_0 & -\mu_0 & -\mu_0 & -\mu_0\\
-\mu_0 & - D_\perp^x k_x^2 - \frac{i k_x v}{2} + 5 \mu_0 & -\mu_0 & -\mu_0 & -\mu_0 & -\mu_0 \\
-\mu_0 & -\mu_0 & - D_\perp^x k_x^2 + \frac{i k_x v}{2} + 5 \mu_0 & -\mu_0 & -\mu_0 & -\mu_0 \\
-\mu_0 & -\mu_0 & -\mu_0 & - D_\parallel^x k_x^2 + i k_x v + 5 \mu_0 & -\mu_0 & -\mu_0 \\
-\mu_0 & -\mu_0 & -\mu_0 & -\mu_0 & - D_\perp^x k_x^2 + \frac{i k_x v}{2} + 5 \mu_0 & -\mu_0 \\
-\mu_0 & -\mu_0 & -\mu_0 & -\mu_0 & -\mu_0 & - D_\perp^x k_x^2 - \frac{i k_x v}{2} + 5 \mu_0 
\end{smallmatrix}\right)
\end{equation}
where $D_\parallel^x = 3D/2(1+\epsilon/5)$ and $D_\perp^x = 3D/2(1-\epsilon/10)$. Up to the order ${\cal O}(k_x^2)$, Mathematica gives the expression of the six eigenvalues $\lambda_{{\rm gas},x}^{i}$:
\begin{gather}
\lambda_{{\rm gas},x}^{1,2} = 6 \mu_0 \pm \frac{iv k_x}{2}-D_\perp^x k_x^2 + \cdots, \\
\lambda_{{\rm gas},x}^{3,4} = 6 \mu_0 \pm \frac{i \sqrt{3} v k_x}{2} - \left( \frac{7 D_\parallel^x+ 2 D_\perp^x}{9} + \frac{v^2}{72 \mu_0} \right) k_x^2 + \cdots, \\
\lambda_{{\rm gas},x}^5 = \left(-\frac{D_\parallel^x+2D_\perp^x}{3} + \frac{v^2}{12\mu_0} \right) k_x^2 + \cdots, \\
\lambda_{{\rm gas},x}^6 = 6\mu_0 - \left(\frac{D_\parallel^x+8D_\perp^x}{9} + \frac{v^2}{18\mu_0} \right) k_x^2+ \cdots.
\end{gather}
Since $D_\parallel^x$ and $D_\perp^x$ are positive the real part of all eigenvalues is negative when $\mu_0<0$. For a perturbation along $y$ ($k_x=0$), the matrix $M_{\rm gas}$ becomes
\begin{equation}
M_{\rm gas}^{(y)}=\left(\begin{smallmatrix}
- D_\parallel^y k_y^2 + 5 \mu_0 & -\mu_0 & -\mu_0 & -\mu_0 & -\mu_0 & -\mu_0\\
-\mu_0 & - D_\perp^y k_y^2 - \frac{i \sqrt{3} k_y v}{2} + 5 \mu_0 & -\mu_0 & -\mu_0 & -\mu_0 & -\mu_0 \\
-\mu_0 & -\mu_0 & - D_\perp^y k_y^2 - \frac{i \sqrt{3} k_y v}{2} + 5 \mu_0 & -\mu_0 & -\mu_0 & -\mu_0 \\
-\mu_0 & -\mu_0 & -\mu_0 & - D_\parallel^y k_y^2 + 5 \mu_0 & -\mu_0 & -\mu_0 \\
-\mu_0 & -\mu_0 & -\mu_0 & -\mu_0 & - D_\perp^y k_y^2 + \frac{i \sqrt{3} k_y v}{2} + 5 \mu_0 & -\mu_0 \\
-\mu_0 & -\mu_0 & -\mu_0 & -\mu_0 & -\mu_0 & - D_\perp^y k_y^2 + \frac{i \sqrt{3} k_y v}{2} + 5 \mu_0 
\end{smallmatrix}\right)
\end{equation}
where $D_\parallel^y = 3D/2(1-\epsilon/5)$ and $D_\perp^y = 3D/2(1+\epsilon/10)$. Up to the order ${\cal O}(k_y^2)$, Mathematica gives the expression of the six eigenvalues $\lambda_{{\rm gas},y}^{i}$:
\begin{gather}
\lambda_{{\rm gas},y}^{1,2} = 6 \mu_0 \pm \frac{i \sqrt{3} v k_y}{2}-D_\perp^y k_y^2 + \cdots, \\
\lambda_{{\rm gas},y}^{3,4} = 6 \mu_0 \pm \frac{i v k_y}{2} - \left( \frac{D_\parallel^y+ 2 D_\perp^y}{3} + \frac{v^2}{24 \mu_0} \right) k_y^2 + \cdots, \\
\lambda_{{\rm gas},y}^5 = \left(-\frac{D_\parallel^y+2D_\perp^y}{3} + \frac{v^2}{12\mu_0} \right) k_y^2 + \cdots, \\
\lambda_{{\rm gas},y}^6 = 6\mu_0 - D_\parallel^y k_y^2.
\end{gather}
Since $D_\parallel^y$ and $D_\perp^y$ are positive we see that the real part of all eigenvalues is negative when $\mu_0<0$.

\subsection{Linear stability of ordered homogeneous solution}
\label{appendixD2}

For an ordered homogeneous solution the ($x$, $y$) symmetry is broken, implying that a perturbation along $x$ axis has a different behavior that a perturbation along $y$ axis. The expression of the matrix $M_{\rm liq}$ is derived from Eq.~(\ref{defMliq}) with $\mu = M (6\beta J - 1 - 2\alpha M + \alpha M^2)$, $\nu = M (6\beta J - 5 + 2\alpha M + 5 \alpha M^2]$ and $\kappa = M [-30 \beta J + 1 +2\alpha M - \alpha M^2]$ for $q=6$. Let us consider first a perturbation in the $x$ direction ($k_y=0$). The matrix then writes
\begin{equation}
M_{\rm liq}^{(x)}=\left(\begin{smallmatrix}
- D_\parallel^x k_x^2 - i k_x v + 5 \mu & \nu & \nu & \nu & \nu & \nu\\
-\mu & - D_\perp^x k_x^2 - \frac{i k_x v}{2} + \kappa & -(\kappa+\nu)/4 & -(\kappa+\nu)/4 & -(\kappa+\nu)/4 & -(\kappa+\nu)/4 \\
-\mu & -(\kappa+\nu)/4 & - D_\perp^x k_x^2 + \frac{i k_x v}{2} + \kappa & -(\kappa+\nu)/4 & -(\kappa+\nu)/4 & -(\kappa+\nu)/4 \\
-\mu & -(\kappa+\nu)/4 & -(\kappa+\nu)/4 & - D_\parallel^x k_x^2 + i k_x v + \kappa & -(\kappa+\nu)/4 & -(\kappa+\nu)/4 \\
-\mu & -(\kappa+\nu)/4 & -(\kappa+\nu)/4 & -(\kappa+\nu)/4 & - D_\perp^x k_x^2 + \frac{i k_x v}{2} + \kappa & -(\kappa+\nu)/4 \\
-\mu & -(\kappa+\nu)/4 & -(\kappa+\nu)/4 & -(\kappa+\nu)/4 & -(\kappa+\nu)/4 & - D_\perp^x k_x^2 - \frac{i k_x v}{2} + \kappa 
\end{smallmatrix}\right)
\end{equation}
where $D_\parallel^x = 3D(1+\epsilon/5)/2$ and $D_\perp^x = 3D(1-\epsilon/10)/2$. Up to the order ${\cal O}(k_x^2)$, Mathematica gives the expression of the six eigenvalues are $\lambda_{{\rm liq}, x}^{i}$:
\begin{gather}
\lambda_{{\rm liq}, x}^{1,2} = \frac{5\kappa+\nu}{4} \pm \frac{ivk_x}{2} -D_\perp^x k_x^2 + \cdots, \\
\lambda_{{\rm liq}, x}^{3,4} = \frac{5\kappa+\nu}{4} \pm \frac{ivk_x}{10}(4+\sqrt{21}) + \left[ \frac{8(D_\parallel^x+D_\perp^x)+\sqrt{21}(2D_\parallel^x+3D_\perp^x)}{5\sqrt{21}} + \frac{8+3\sqrt{21}}{125\sqrt{21}}\frac{6(5\kappa-20\mu+\nu)v^2}{(5\kappa+\nu)(\kappa-4\mu+\nu)}\right] k_x^2 + \cdots, \\
\lambda_{{\rm liq}, x}^{5} = (5\mu-\nu)+ \frac{i(25\mu+\nu)vk_x}{5(5\mu-\nu)} + \left[\frac{(4D_\perp^x+D_\parallel^x)\nu-25D_\parallel^x \mu}{5(5\mu-\nu)}-\frac{36\nu[-25(\kappa-5\mu)\mu -35 \mu \nu + \nu^2]v^2}{125(5\mu-\nu)^3(\kappa-4\mu+\nu)}\right] k_x^2 + \cdots, \\
\lambda_{{\rm liq}, x}^{6} = \frac{i(\mu+\nu)vk_x}{5\mu-\nu} + \left[\frac{-(D_\parallel^x+4D_\perp^x)\mu + D_\parallel^x \nu}{5\mu-\nu} - \frac{36\mu(-5\mu^2+2\mu\nu+\kappa\nu)v^2}{(5\mu-\nu)^3(5\kappa+\nu)}\right] k_x^2 + \cdots.
\end{gather}

Now we look at a perturbation in the $y$ direction ($k_x=0$). The matrix $M_{\rm liq}$ writes then
\begin{equation}
M_{\rm liq}^{y}=\left(\begin{smallmatrix}
- D_\parallel^y k_y^2 + 5 \mu & \nu & \nu & \nu & \nu & \nu\\
-\mu & - D_\perp^y k_y^2 - \frac{i \sqrt{3} k_y v}{2} + \kappa & -(\kappa+\nu)/4 & -(\kappa+\nu)/4 & -(\kappa+\nu)/4 & -(\kappa+\nu)/4 \\
-\mu & -(\kappa+\nu)/4 & - D_\perp^y k_y^2 - \frac{i \sqrt{3} k_y v}{2} + \kappa & -(\kappa+\nu)/4 & -(\kappa+\nu)/4 & -(\kappa+\nu)/4 \\
-\mu & -(\kappa+\nu)/4 & -(\kappa+\nu)/4 & - D_\parallel^y k_y^2 + \kappa & -(\kappa+\nu)/4 & -(\kappa+\nu)/4 \\
-\mu & -(\kappa+\nu)/4 & -(\kappa+\nu)/4 & -(\kappa+\nu)/4 & - D_\perp^y k_y^2 + \frac{i \sqrt{3} k_y v}{2} + \kappa & -(\kappa+\nu)/4 \\
-\mu & -(\kappa+\nu)/4 & -(\kappa+\nu)/4 & -(\kappa+\nu)/4 & -(\kappa+\nu)/4 & - D_\perp^y k_y^2 + \frac{i \sqrt{3} k_y v}{2} + \kappa 
\end{smallmatrix}\right)
\end{equation}
where $D_\parallel^y = 3D(1-\epsilon/5)/2$ and $D_\perp^y = 3D(1+\epsilon/10)/2$. Up to the order ${\cal O}(k_y^2)$, Mathematica gives the expression of the six eigenvalues are $\lambda_{{\rm liq}, y}^{i}$:
\begin{gather}
\lambda_{{\rm liq}, y}^{1,2} = \frac{5\kappa+\nu}{4} \pm \frac{i\sqrt{3}vk_y}{2} -D_\perp^y k_y^2 + \cdots, \\
\lambda_{{\rm liq}, y}^{3,4} = \frac{5\kappa+\nu}{4} \pm \frac{i\sqrt{3/5}vk_y}{2} + \left[\frac{-2D_\parallel^y+3D_\perp^y}{5} - \frac{6(5\kappa-20\mu+\nu)v^2}{25(5\kappa + \nu)(\kappa-4\mu+\nu))}\right] k_y^2 + \cdots, \\
\lambda_{{\rm liq}, y}^{5} = (5\mu-\nu) + \left[\frac{(4D_\perp^y+D_\parallel^y)\nu-25D_\parallel^y \mu}{5(5\mu-\nu)}-\frac{12\nu v^2}{25(5\mu-\nu)(\kappa-4\mu+\nu)}\right]k_y^2 + \cdots, \\
\lambda_{{\rm liq}, y}^{6} = \left[\frac{-(D_\parallel^y+4D_\perp^y)\mu + D_\parallel^y \nu}{5\mu-\nu} + \frac{12\mu v^2}{(5\mu-\nu)(5\kappa+\nu)}\right]k_y^2 + \cdots.
\end{gather}

The ordered homogeneous solution is then stable if $5\kappa+\nu<0$ and $5\mu-\nu<0$ for the two different perturbations. Since $5\mu - \nu = 12M ( 2\beta J - \alpha M)$, the only stable ordered homogeneous solution satisfies $M> 2 \beta J/ \alpha$. From Eq.~(\ref{eqMag2}), the magnetization of the stable solution is then equal to
\begin{equation}
\label{stableM6}
M = \frac{2\beta J}{\alpha} + \sqrt{\frac{r}{\alpha \rho_*}} \sqrt{\frac{\rho_0-\rho_*}{\rho_0}} = M_0 + M_1 \delta,
\end{equation}
where $\rho_*$ has been defined in Eq.~(\ref{defrhostar}); $M_0=2\beta J / \alpha$ and $M_1 = \sqrt{r/\alpha \rho_*}$ are temperature dependent constants; and $\delta = \sqrt{(\rho_0-\rho_*)/\rho_0}$ is a variable with values between $0$ and $1$. Moreover, $5\kappa + \nu = 12M(-12 \beta J + \alpha M)$ implying that $M< 12 \beta J / \alpha$ to have a stable solution, which is always satisfied from the maximal value of M: $M<3\beta J/\alpha$ from Eq.~(\ref{stableM6}).

However, the stability of the two different perturbations differs from $\lambda_{{\rm liq}, x}^6$ and $\lambda_{{\rm liq}, y}^6$. The perturbation along $x$ is stable only if
\begin{equation}
\label{lambdaPara6app}
\lambda_\parallel = \frac{\Re \lambda_{{\rm liq}, x}^6}{k_x^2} = -D + \frac{\mu+\nu}{5\mu-\nu} \frac{3D\epsilon}{10} - \frac{36\mu(-5\mu^2+2\mu\nu+\kappa\nu)}{(5\mu-\nu)^3(5\kappa+\nu)} \left(\frac{6D\epsilon}{5} \right)^2
\end{equation}
is negative and the $y$-perturbation is stable if
\begin{equation}
\label{lambdaPerp6app}
\lambda_\perp = \frac{\Re \lambda_{{\rm liq}, y}^6}{k_y^2} = -D - \frac{\mu+\nu}{5\mu-\nu} \frac{3D\epsilon}{10} + \frac{12\mu}{(5\mu-\nu)(5\kappa+\nu)} \left(\frac{6D\epsilon}{5} \right)^2
\end{equation}
is negative. These two eigenvalues can be rewritten as
\begin{gather}
\label{defEIGEN6}
\lambda_\parallel = -D + Q_1 \frac{3D\epsilon}{10} + Q_3 \left(\frac{6D\epsilon}{5} \right)^2, \qquad
\lambda_\perp = -D - Q_1 \frac{3D\epsilon}{10} + Q_2 \left(\frac{6D\epsilon}{5} \right)^2
\end{gather}
for the quantities $Q_i$ independent of $\epsilon$ defined by
\begin{equation}
Q_1 = \frac{\mu+\nu}{5\mu-\nu}, \qquad Q_2 = \frac{12\mu}{(5\mu-\nu)(5\kappa+\nu)}, \qquad Q_3 = - \frac{36\mu(-5\mu^2+2\mu\nu+\kappa\nu)}{(5\mu-\nu)^3(5\kappa+\nu)}.
\end{equation}
From this stability analysis we can infer that transverse bands will be formed when $\lambda_\perp < 0 < \lambda_\parallel$ whereas longitudinal lanes will be created in the opposite case $\lambda_\parallel < 0 < \lambda_\perp$. Then, the reorientation transition happens when $\lambda_\parallel = 0$ and $\lambda_\perp =0$, and from the Eq.~(\ref{defEIGEN6}) the value of the drift $\epsilon_*$ at the reorientation transition is given by
\begin{equation}
\label{eqepsilonstar6}
\frac{6D\epsilon_*}{5} = \sqrt{\frac{3D}{Q_2+Q_3}} = \frac{-Q_1}{2(Q_3-Q_2)},
\end{equation}
where the second equality defines the value of $\rho_0$ where the reorientation transition takes place, which can be rewritten as
\begin{equation}
\label{eqdeltastar6}
-Q_1 = \frac{2\sqrt{3D}(Q_3-Q_2)}{\sqrt{Q_2+Q_3}}.
\end{equation}

With Eq.~(\ref{stableM6}) we can now rewrite the quantities $Q_1$, $Q_2$ and $Q_3$ as a function of $\delta$ with coefficients depending only on $M_0$ and $M_1$. With Mathematica, we obtain after simplifications
\begin{gather}
Q_1(\delta) = -M_0 - \frac{M_1}{2} \frac{1+\delta^2}{\delta},\\
Q_2(\delta) = \frac{M_0^2+M_0-M_1^2}{(-5 M_0 + \delta M_1)(M_0 + \delta M_1)} \left[ - \frac{M_1}{12} \frac{1+\delta^2}{\delta} + \frac{1-M_0}{6}\right], \\
Q_3(\delta) = -Q_2(\delta) \left[ \frac{M_1}{4} \frac{1+\delta^2}{\delta} - \frac{5M_0}{4 \delta^2} - \frac{2+3 M_0}{4} - \frac{M_0(5M_0+1)}{2\delta M_1} \right].
\end{gather}
The Eq.~(\ref{eqdeltastar6}) is then satisfied for $\delta=\delta_*$. Since the inversion of this equation to get the exact expression of $\delta_*$ is too complicated, we look at the solution close to the critical point: $T \rightarrow T_c$, similarly to the $q=4$ case. For this limiting case, the ordered-disordered transition takes place for $\rho_* \gg 1$, implying that $M_1 \ll 1$. We can then look at a solution $\delta_*$ as an asymptotic expansion in $M_1$, such that $\delta_* = \delta_1 M_1 + \delta_2 M_1^2 + \delta_3 M_1^3 + \cdots$. Taking $D=1$, we get with Eq.~(\ref{eqdeltastar6})
\begin{equation}
\delta_* = \frac{1}{2(1-M_0)} M_1 - \frac{2-3M_0}{24(M_0-1)^3(1+M_0)} M_1^3 + {\cal O}(M_1^5)
\end{equation}
and the quantities $Q_1$, $Q_2$ and $Q_3$ evaluated at $\delta= \delta_*$ are equal to
\begin{gather}
Q_1(\delta_*) = -1 + \frac{5 M_1^2}{12(-1+M_0^2)} + {\cal O}(M_1^4),\\
Q_2(\delta_*) = \frac{M_1^2}{72 M_0 (1-M_0)} + {\cal O}(M_1^4), \\
Q_3(\delta_*) = \frac{1}{12} - \frac{(3+8M_0)M_1^2}{72 M_0(1-M_0^2)} + {\cal O}(M_1^4).
\end{gather}
Thus, with Eq.~(\ref{eqepsilonstar6}), we obtain the expression of $\epsilon_*$ where the reorientation transition occurs:
\begin{equation}
\epsilon_* = 5 \left[1- \frac{4+9M_0}{12 M_0(1-M_0^2)} M_1^2 + {\cal O}(M_1^4) \right].
\end{equation}
In section \ref{section4}, we have shown that $T_c^{-1} = 1 - \sqrt{5/12}$. Then we get
\begin{equation}
M_1^2 \simeq \frac{2}{7} (-10+3\sqrt{15}) \frac{T_c-T}{T_c}, \qquad M_0 = \frac{1}{7} (-1+\sqrt{15}),
\end{equation}
leading to the expression of $\epsilon_*$ as
\begin{equation}
\epsilon_* = 5 \left[1 + \frac{150-67\sqrt{15}}{126} \frac{T_c-T}{T_c} + \cdots \right] \simeq 5 \left[1 - 0.869 \frac{T_c-T}{T_c} + \cdots \right] \simeq \frac{5T}{T_c} + 0.655 \frac{T_c-T}{T_c} + \cdots.
\end{equation}
So, to leading order in $T_c-T$ the reorientation transition occurs for $\epsilon_* = 5 T/T_c$.

\bibliography{PostDoc.bib}

\end{document}